\documentclass[aps, a4paper, floatfix, nofootinbib, longbibliography, twocolumn]{revtex4-1}

\usepackage[utf8]{inputenc}
\usepackage{epsfig}
\usepackage[T1]{fontenc}
\usepackage[english]{babel}
\usepackage[table]{xcolor}
\usepackage{t1enc}
\usepackage{graphicx}
\usepackage{amssymb}
\usepackage{amsmath}
\usepackage{relsize}
\usepackage{overpic}
\usepackage{bm}
\usepackage{hyperref}
\usepackage{times}
\usepackage[section]{placeins}

\usepackage[normalem]{ulem}

\newcommand{\dd}{\mathrm{d}}

\usepackage{mathtools}
\def\multiset#1#2{\ensuremath{\left(\kern-.3em\left(\genfrac{}{}{0pt}{}{#1}{#2}\right)\kern-.3em\right)}}

\usepackage{amsmath}

\usepackage{verbatim}
\usepackage{overpic}
\usepackage{booktabs}
\usepackage{placeins}

\newcommand{\p}{\bm{p}}
\newcommand{\x}{\bm{x}}
\renewcommand{\ss}{\bm{s}}
\newcommand{\lb}{\bm{l}}

\begin{document}

\title{Change points, memory and epidemic spreading in temporal networks}

\author{Tiago P. Peixoto}
\affiliation{Department of Mathematical Sciences and Centre for Networks
and Collective Behaviour, University of Bath, Claverton Down, Bath BA2
7AY, United Kingdom}
\affiliation{ISI Foundation, Via Chisola 5, 10126 Torino, Italy}
\author{Laetitia Gauvin}
\affiliation{ISI Foundation, Via Chisola 5, 10126 Torino, Italy}


\begin{abstract}
  Dynamic networks exhibit temporal patterns that vary across different
  time scales, all of which can potentially affect processes that take
  place on the network. However, most data-driven approaches used to
  model time-varying networks attempt to capture only a single
  characteristic time scale in isolation --- typically associated with
  the short-time memory of a Markov chain or with long-time abrupt
  changes caused by external or systemic events. Here we propose a
  unified approach to model both aspects simultaneously, detecting
  short and long-time behaviors of temporal networks. We do so by
  developing an arbitrary-order mixed Markov model with change points,
  and using a nonparametric Bayesian formulation that allows the Markov
  order and the position of change points to be determined from data
  without overfitting. In addition, we evaluate the quality of the
  multiscale model in its capacity to reproduce the spreading of
  epidemics on the temporal network, and we show that describing
  multiple time scales simultaneously has a synergistic effect, where
  statistically significant features are uncovered that otherwise would
  remain hidden by treating each time scale independently.
\end{abstract}

\maketitle

\section{Introduction}

Recent advances in the study of network systems --- usually with social,
technological and biological origins --- have been moving beyond the
more traditional approach of considering them as static or growing
entities, and instead have been introducing more realistic descriptions
that allow them to change arbitrarily in
time~\cite{holme_temporal_2012,holme_modern_2015}. This effort includes
modeling of the time-varying network structure~\cite{ho2011evolving,perra2012activity}, as well as processes
that take place on this dynamic environment, such as epidemic
spreading~\cite{rocha_simulated_2011, valdano_analytical_2015,
  genois_compensating_2015,ren2014epidemic}. Further recent
works~\cite{karsai2011small,gauvin2013activity,vestergaard2014memory}
have highlighted the role of memory, burstiness and time ordering as key
features of empirical temporal networks that affect dynamical processes
taking place on it.

Most approaches, however, rely on a
characteristic time scale on which they describe the dynamics. These can
be divided, roughly, into approaches that model temporal correlations
via Markov chains relating short-time memory with future
behavior~\cite{scholtes_causality-driven_2014,peixoto_modelling_2017},
and those that model the dynamics at longer times, usually via network
snapshots ~\cite{xu_dynamic_2013,gauvin_detecting_2014,peixoto_inferring_2015,
  stanley_clustering_2016,ghasemian_detectability_2016,zhang_random_2017}
or discrete change points~\cite{peel_detecting_2015,de2016detection,cornelimultiple}. In reality,
however, most systems exhibit both kinds of dynamics, and focusing on a
single aspect comes at the expense of ignoring the other. In this work,
we introduce a data-driven modeling approach that includes both aspects
simultaneously, and is capable of uncovering both the short-time Markov
properties as well a the long-time abrupt changes.

We develop a Bayesian formulation that allows both the change points and
the Markov order to be inferred from data in a principled manner,
prevents overfitting and enables model selection. As an extraneous
evaluation of our approach, we investigate the behavior of epidemic
spreading both in the original data and in artificial ones generated
from our inferred models. We show that the most plausible models tend to
mix both short-time memory and many change points, and those tend to
capture well the nontrivial epidemic behavior observed in the original
data. Importantly, the inferred models with change points typically
uncover higher-order memory than the simpler stationary variants,
demonstrating that the mixed approach is more powerful than considering
individual ones in isolation.

This paper is divided as follows. In Sec.~\ref{sec:epidemics} we present
the epidemic models that will be used for the model comparison. In
Sec.~\ref{sec:model} we describe our modeling and inferring approach,
and apply it to empirical data.  In Sec.~\ref{sec:conclusion} we
finalize with a conclusion.

\section{Results}

\subsection{Proximity networks and epidemic dynamics}\label{sec:epidemics}

In the interest of simplicity, we will consider a minimal model of
temporal networks and epidemic dynamics that takes place on it. The
most central simplification we will make is that the dynamics takes
place in discrete time, so that the placement of edges forms a temporal
sequence, where only one edge is placed at any given time. Real
dynamical networks and epidemic spreading occur in continuous time, but
our objective here is not to construct a detailed realistic model, but
rather to illustrate how multiple time scales can be described
simultaneously. More realistic features can then be added to the model
at a later stage.

More specifically, we consider temporal networks that occur as a
sequence of edges $\ss=\{x_t\}$, where $x_t=(u,v)_t$ is an edge between
nodes $u$ and $v$ observed at time $t$, with $t=\{1,2,\dots,E\}$ where
$E$ is the number of edges. Although this formulation is general, we
focus in particular on proximity networks, obtained by tracking
volunteers with wearable sensors over a period of
time~\cite{toroczkai_proximity_2007,
stehle_high-resolution_2011,vanhems_estimating_2013,mastrandrea_contact_2015},
so that an edge $(u,v)_t$ is recorded if the respective people came
closer than a given radius at time $t$. Data recorded in this manner
possess enough time resolution for our analysis, and also serve as a
plausible scenario for epidemic
spreading~\cite{gemmetto_mitigation_2014}.

In the above scenario, we assume that an infection can only occur at
time $t$ over the current ``active'' edge $(u,v)_t$. If the epidemics
follows the Susceptible-Infected-Recovered (SIR) model, and $\sigma_u(t)
\in \{\text{S},\text{I},\text{R}\}$ is the state of node $u$ at time $t$,
we have at each time step $t$:
\begin{enumerate}
\item If $(u, v)_t$ is the current edge, with $(\sigma_u(t-1),
  \sigma_v(t-1)) = (S, I)$ or $(I, S)$, the infection spreads with
  probability $\beta$, so that $(\sigma_u(t), \sigma_v(t)) = (I,
  I)$.
\item For every infected node $u$ with $\sigma_u(t-1)=I$, it becomes
      recovered $\sigma_u(t)=R$ with probability $\gamma$. 
\end{enumerate}
The parameters $\beta$ and $\gamma$ control the infection and recovery
rates, respectively. We also consider the
Susceptible-Infected-Susceptible (SIS) model, which is a variation of
the above, where in the second step the infected nodes become susceptible,
$\sigma_u(t)=S$, instead of recovered. In both cases, we consider the
total number of infected nodes at given time $t$,
\begin{equation}
  X(t) = \sum_u\delta_{\sigma_u(t),I}
\end{equation}
where $\delta$ is the Kronecker delta.
For any positive recovery rate $\gamma > 0$, the long-time behavior of
the SIR model is always $\lim_{t\to\infty}X(t) = 0$, as the outbreak
invariably dies out, whereas in the SIS model it can have an indefinite
permanence. In the following, we will use the behavior of $X(t)$ as a
proxy for the comparison between data and model in capturing the
underlying network dynamics.

When considering epidemics on dynamical networks, there are two
properties that are believed to be crucial for the spreading
process~\cite{gauvin2013activity,vestergaard2014memory}: 1. The distribution of number of contacts per link,
i.e. the frequency of token $x$ in sequence $\ss$, and 2. The
distribution of waiting (or inter-event) times, i.e. the time between
two occurrences of the same edge. We will have these two aspects in
mind when elaborating our models.

\subsection{Models for temporal networks}\label{sec:model}

Our objective is to construct a generative model for temporal networks
that includes both short-term memories and abrupt change points. We
begin by formulating a stationary version, without change points, and
show how it is insufficient to capture many features in the data. We
then extend the model to include change points, and perform a
comparison.

\subsubsection{Stationary Markov chains}
We consider sequences of discrete tokens, i.e. edges, $\ss=\{x_t\}$ with
$t\in \{1,\dots,E\}$ being the time and $x_t\in\{1,\dots,D\}$ the
set of unique edges with cardinality $D$, which are generated from a
stationary Markov chain of order $n$, i.e. they occur with probability
\begin{equation}\label{eq:markov}
  P(\ss|\p,n) = \prod_t p_{x_t,\x_{t-1}} = \prod_{x,\x}p_{x,\x}^{a_{x,\x}},
\end{equation}
where $\p$ corresponds to the transition matrix and $p_{x_t,\x_{t-1}}$
is the probability of observing token $x_t$ given the previous $n$
tokens $\x_{t-1}=\{x_{t-1},\dots,x_{t-n}\}$ in the sequence, and
$a_{x,\x}$ is the number of observed transitions from memory $\x$ to
token $x$. This serves a simple model for temporal networks, where each
possible token corresponds to an edge in the network, i.e. $x_t \equiv
(i,j)_t$, as we considered previously. Despite its simplicity, this
model is able to reproduce arbitrary edge frequencies, determined by the
steady-state distribution of the tokens $x$, as well as causal temporal
correlations between edges. This means that the model should be able to
reproduce properties of the data that can be attributed to the distribution
of number of contacts per link, which are believed to be important for
epidemic spreading~\cite{gauvin2013activity,vestergaard2014memory}. However, due to its Markovian nature, the
dynamics will eventually forget past states, and converge to the
limiting distribution (assuming the chain is ergodic and
aperiodic). This latter property means that the model should be able to
capture nontrivial statistics of waiting times only at a short time
scale, comparable to the Markov order.

\begin{figure}
  \includegraphics[width=\columnwidth]{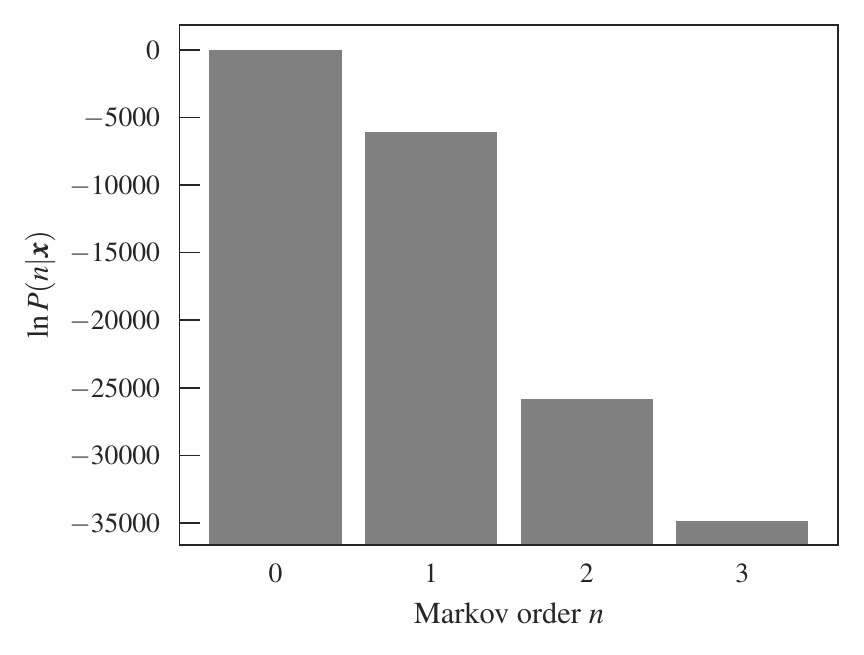}
  \caption{Posterior
  distribution of the Markov order $P(n|\x)$ (Eq.~\ref{eq:markov_posterior}) for a temporal network
  between students in a high
  school~\cite{fournet_contact_2014}.\label{fig:markov-order}}
\end{figure}
\begin{figure}
  \begin{overpic}[width=\columnwidth]{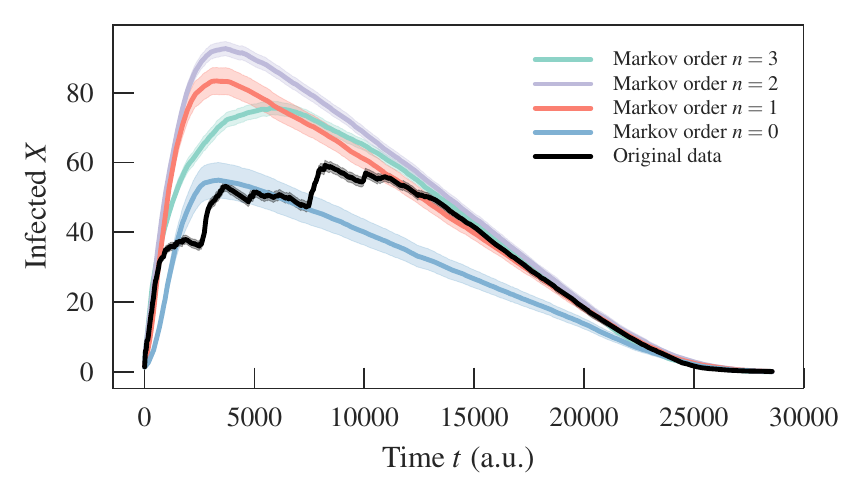}\put(0,5){(a)}\end{overpic}\\
  \begin{overpic}[width=\columnwidth]{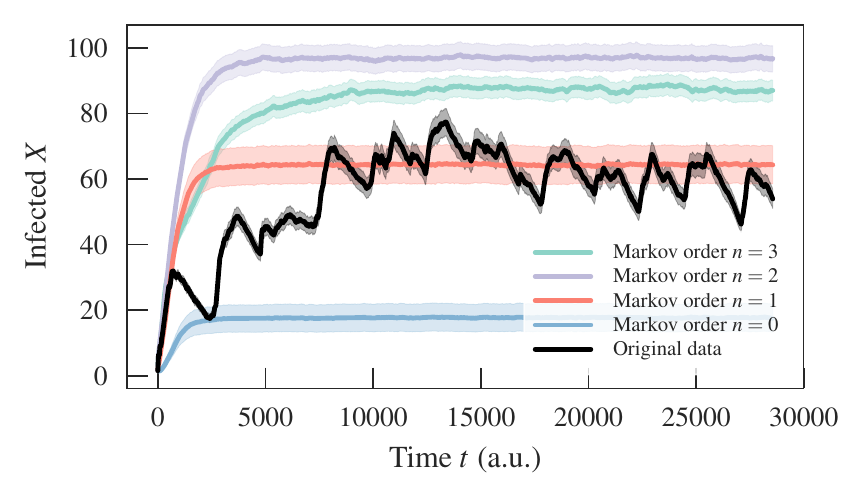}\put(0,5){(b)}\end{overpic}
  \caption{Number of infected nodes over time $X(t)$ for a temporal
  network between students in a high-school~\cite{fournet_contact_2014},
  considering both the original data and artificial time-series
  generated from the fitted Markov model of a given order $n$, using (a)
  SIR ($\beta=0.41$, $\gamma=0.005$) and (b) SIS
  ($\beta=0.61$, $\gamma=0.03$) epidemic models. In all cases, the values
  were averaged over 100 independent realizations of the network model
  (for the artificial datasets) and dynamics. The shaded areas indicate
  the standard deviation of the mean. The values of the infection and recovery rates were chosen so that the spreading dynamics spans the entire time range of the dataset. \label{fig:markov-epidemics} }
\end{figure}

Given the above model, the simplest way to proceed would be to infer
transition probabilities from data using maximum likelihood,
i.e. maximizing Eq.~\ref{eq:markov} under the normalization constraint
$\textstyle\sum_xp_{x,\x}=1$. This yields
\begin{equation}
  \hat{p}_{x,\x} = \frac{a_{x,\x}}{k_{\x}}.
\end{equation}
where $k_{\x}=\sum_xa_{x,\x}$ is the number of transitions originating
from $\x$.
However, if we want to determine the most appropriate Markov order
$n$ that fits the data, the maximum likelihood approach cannot be used,
as it will \emph{overfit}, i.e. the likelihood of Eq.~\ref{eq:markov}
will increase monotonically with $n$, favoring the most complicated
model possible, and thus confounding statistical fluctuations with
actual structure. Instead, the most appropriate way to proceed is to
consider the Bayesian posterior distribution
\begin{equation}\label{eq:markov_posterior}
  P(n|\ss) = \frac{P(\ss|n)P(n)}{P(\ss)},
\end{equation}
which involves the integrated marginal
likelihood~\cite{strelioff_inferring_2007}
\begin{equation}\label{eq:markov_int}
  P(\ss|n) = \int P(\ss|\p,n)P(\p|n)\,\dd\p,
\end{equation}
where the prior probability $P(\p|n)$ encodes the amount of knowledge we
have on the transitions $\p$ before we observe the data. If we possess no
information, we can be agnostic by choosing a uniform prior
\begin{equation}\label{eq:p_prior}
  P(\p|n) = \prod_{\x}(D-1)!\delta\left(1-\textstyle\sum_xp_{x,\x}\right),
\end{equation}
which assumes that all transition probabilities are equally likely.
Inserting Eq.~\ref{eq:markov} and~\ref{eq:p_prior} in
Eq~\ref{eq:markov_int}, and calculating the integral we obtain
\begin{equation} \label{eq:markov_conditional_order}
  P(\ss|n) = \prod_{\x}\frac{(D-1)!}{(k_{\x}+D-1)!}\prod_x a_{x,\x}!.
\end{equation}
The remaining prior, $P(n)$, that represents our \emph{a
priori} preference to the Markov order, can also be chosen in an
agnostic fashion in a range $[0,N]$, i.e.
\begin{equation}\label{eq:markov_order}
  P(n) = \frac{1}{N+1}.
\end{equation}
Since this prior is a constant, the upper bound $N$ has no effect on the
posterior of Eq.~\ref{eq:markov_posterior}, provided it is
sufficiently large to include most of the distribution.

Differently from the maximum-likelihood approach described previously,
the posterior distribution of Eq.~\ref{eq:markov_posterior} will select
the size of the model to match the statistical significance available,
and will favor a more complicated model only if the data cannot be
suitably explained by a simpler one, i.e. it corresponds to an
implementation of Occam's razor that prevents overfitting.

When applying this approach to empirical data, we observe that it favors
$n=0$ for all datasets we considered (not shown), indicating that a
higher-order model is not statistically justified, as can be seen in
Fig.~\ref{fig:markov-order}.  However, if we generate temporal networks
from the fitted models,
i.e. sequence of edges using the transition probabilities
$\hat{p}_{x,\x} = a_{x,\x}/k_{\x}$, they exhibit epidemic dynamics that
are very different from what we observe on the empirical time-series, as
can be seen in Fig.~\ref{fig:markov-epidemics}: for the original data,
the epidemic spreading is marked by abrupt changes in the infection
rate, which are not reproduced by the model for any value of Markov
order $n$ --- even those that overfit. Therefore, these patterns in the
epidemic dynamics seem to stem from changes in the underlying structure
of the temporal network that are not captured by the above Markov
model. Among other things, this means that the behavior cannot be
explained by a heterogeneous distribution of edge frequencies, as this
is well described by the model. As we show in the next section, the
situation changes considerably once we generalize the model to
incorporate heterogeneous Markov chains with change points.

\subsubsection{Markov chains with change points}

\begin{figure*}
    \begin{minipage}{\textwidth}\centering
      \begin{overpic}[width=.49\columnwidth]{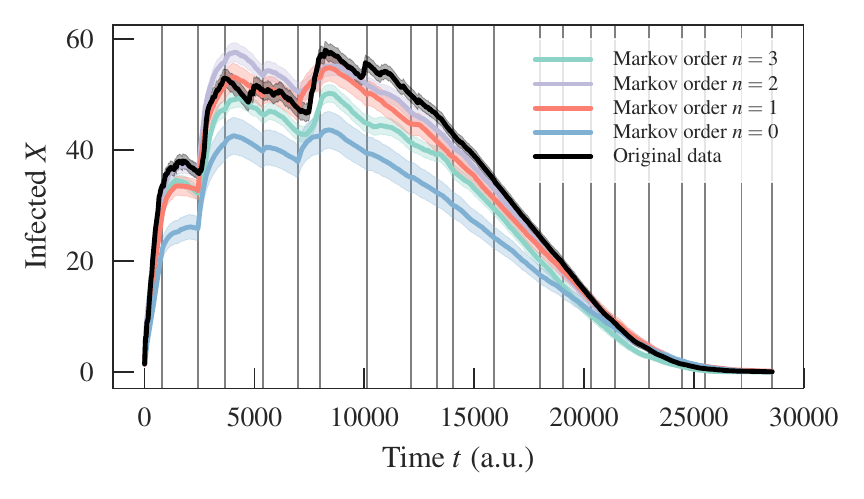}\put(0,5){(a)}\end{overpic}
      \begin{overpic}[width=.49\columnwidth]{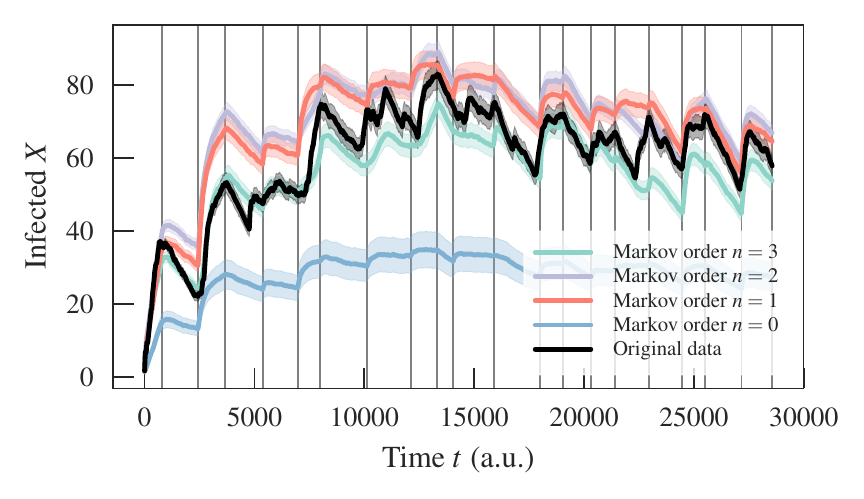}\put(0,5){(b)}\end{overpic}\\
      \includegraphics[width=.19\columnwidth]{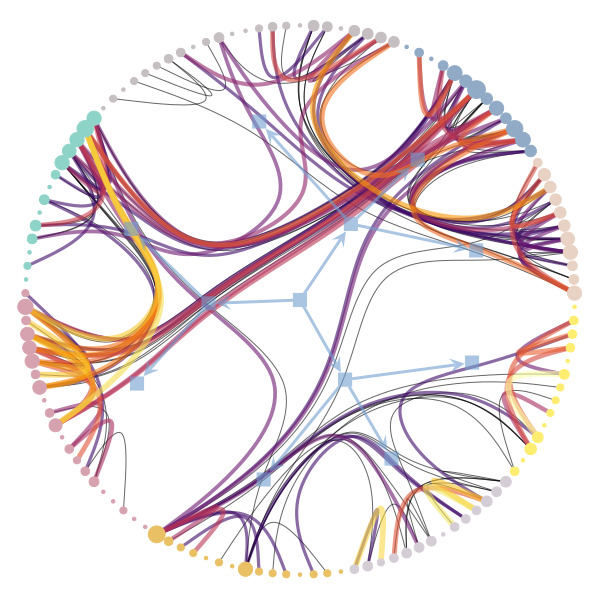}
      \includegraphics[width=.19\columnwidth]{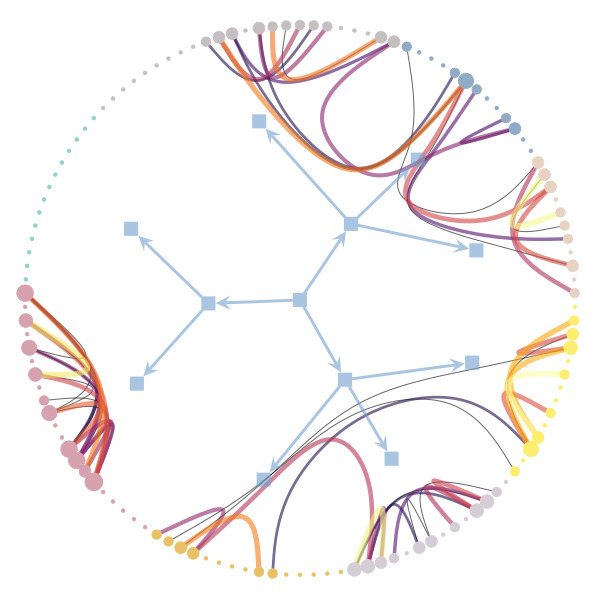}
      \includegraphics[width=.19\columnwidth]{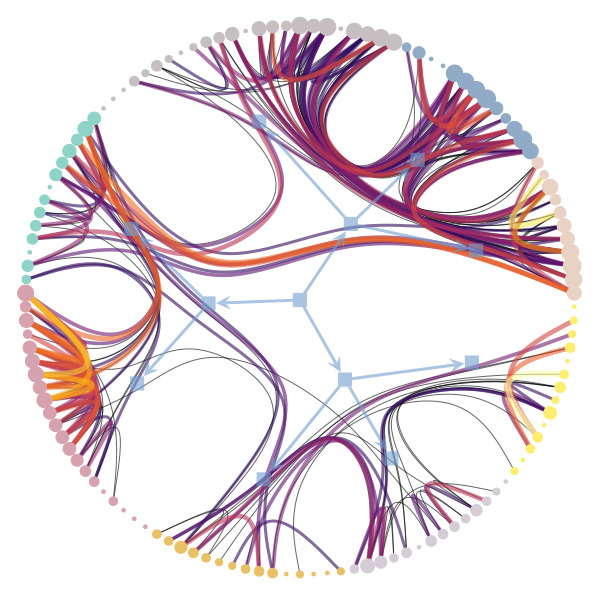}
      \includegraphics[width=.19\columnwidth]{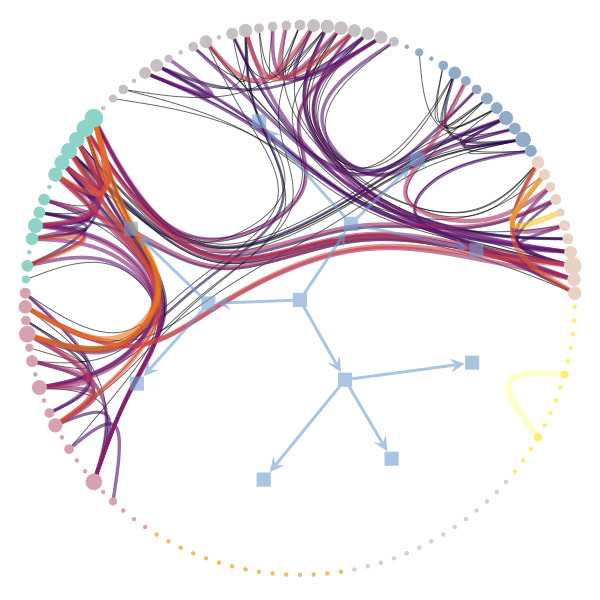}
      \includegraphics[width=.19\columnwidth]{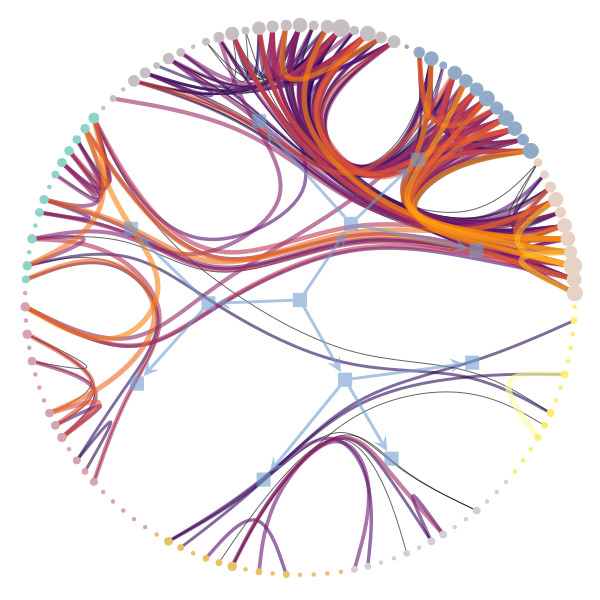}\\
      \includegraphics[width=.19\columnwidth]{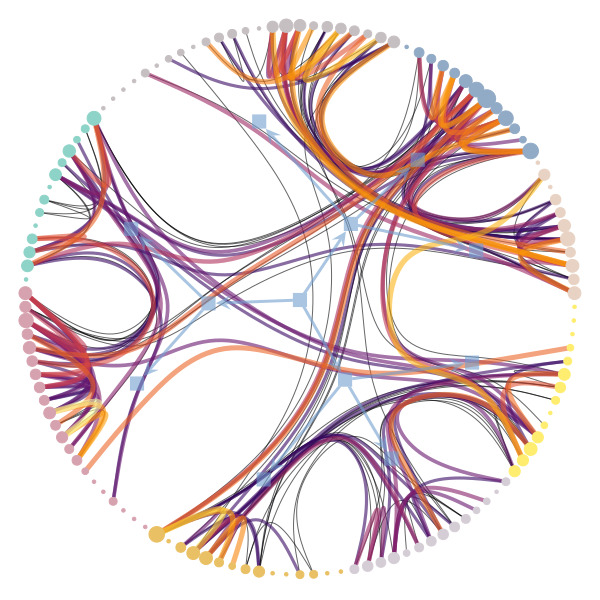}
      \includegraphics[width=.19\columnwidth]{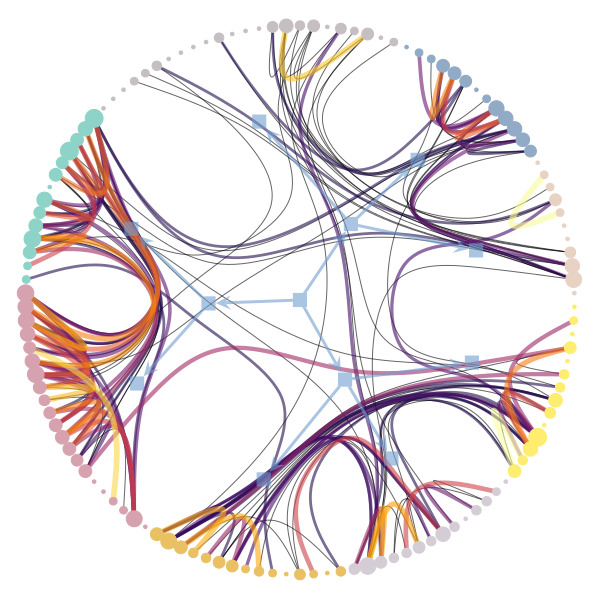}
      \includegraphics[width=.19\columnwidth]{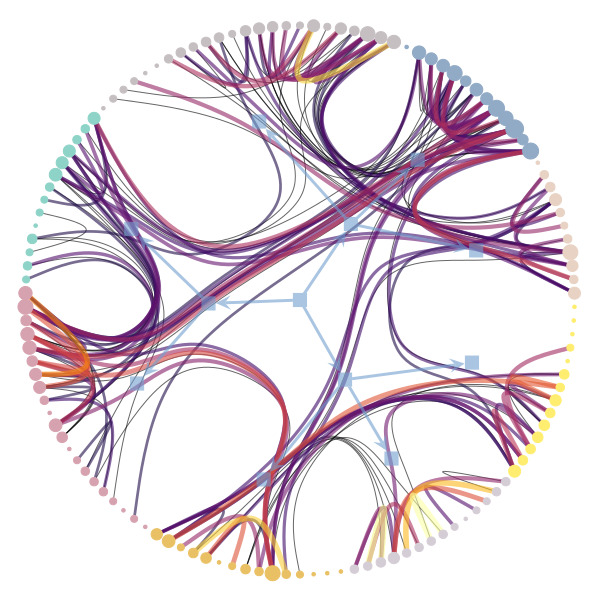}
      \includegraphics[width=.19\columnwidth]{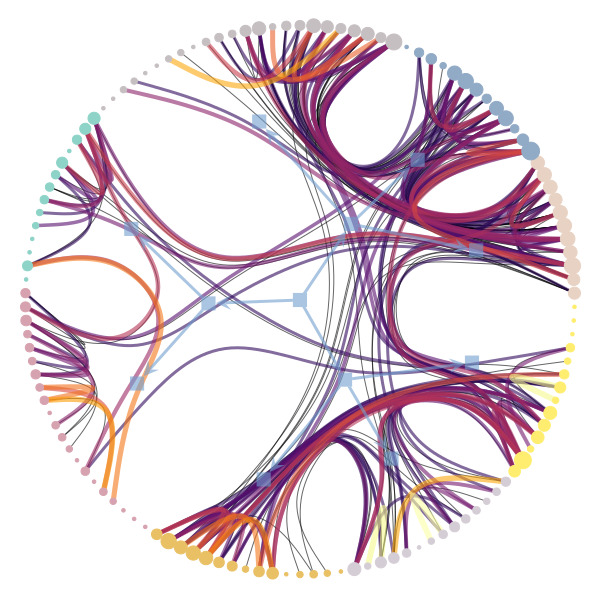}
      \includegraphics[width=.19\columnwidth]{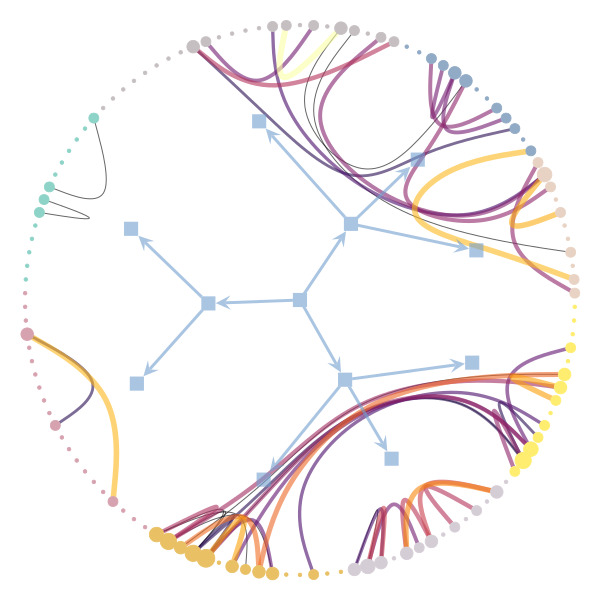}
    \end{minipage}
    \caption{(Above) Number of infected nodes over time $X(t)$ for a temporal
  network between students in a high-school~\cite{fournet_contact_2014},
  considering both the original data and artificial time-series
  generated from the fitted nonstationary Markov model of a given order
  $n$, using (a) SIR ($\beta=0.41$, $\gamma=0.005$) and (b) SIS
  ($\beta=0.61$, $\gamma=0.03$) epidemic models. The vertical lines mark
  the position of the inferred change points. In all cases, the values
  were averaged over 100 independent realizations of the network model
  (for the artificial datasets) and dynamics. The shaded areas indicate
  the standard deviation of the mean. (Below) Network structure inside
  the first ten segments, as captured by a layered hierarchical
  degree-corrected stochastic block model~\cite{peixoto_inferring_2015}
  using the frequency of interactions as edge
  covariates~\cite{peixoto_nonparametric_2017} (indicated by colors),
  where each segment is considered as a different layer. The values of
  the infection and recovery rates were chosen so that the spreading
  dynamics spans the entire time range of the
  dataset.\label{fig:markov-changepoints}}
\end{figure*}

We attempt to model the abrupt changes observed in the previous section
by non-stationary transition probabilities $p_{x,\vec{x}}$ that change
abruptly at a given ``change point,'' but otherwise remain constant
between change points. The occurrence of change points is governed by
the probability $q$ that one is inserted at any given time. The
existence of $M$ change points divide the time series into $M+1$
temporal segments indexed by $l\in\{0,\dots,M\}$. The variable $l_t$
indicates to which temporal segment a given time $t$ belongs among the
$M$ segments. Thus, the conditional probability of observing a token $x$
at time $t$ in segment $l_t$ is given by
\begin{equation}
  P(x_t, l_t|\x_{t-1},l_{t-1}) = p_{x,\x}^{l_t}[q(1-\delta_{l_t,l_{t-1}})+(1-q)\delta_{l_t,l_{t-1}}],
\end{equation}
where $p_{x,\x}^{l_t}$ is the transition probability inside segment $l_t$ and
$q$ is the probability to transit from segment $l$ to $l+1$.  The
probability of a whole sequence $\ss=\{x_t\}$ and $\lb=\{l_t\}$ being
generated is then
\begin{equation}
  P(\ss, \lb|\p,q) = q^M(1-q)^{E-M}\prod_{l,x,\x}\left(p_{x,\x}^{l}\right)^{a_{x,\x}^{l}}
\end{equation}
where $a_{x,\x}^{l}$ is the number of transitions from
memory $\x$ to token $x$ in the segment $l$.
Note that we recover the stationary model of Eq.~\ref{eq:markov} by
setting $q=0$. The maximum-likelihood estimates of the parameters are
\begin{equation}
  \hat{p}_{x,\x}^{l} = \frac{a_{x,\x}^l}{k_{\x}^l}, \quad \hat{q} = \frac{M}{E}
\end{equation}
where $k_{\x}^l=\sum_xa_{x,\x}^{l}$ is the number of transitions originating
from $\x$ in a segment $l$.
But once more, we want to infer the model the segments $\lb$ in a
Bayesian way, via the posterior distribution
\begin{equation}\label{eq:posterior_cp}
  P(\lb|\ss,n) = \frac{P(\ss, \lb|n)}{P(\ss|n)},
\end{equation}
where the numerator is the integrated likelihood
\begin{equation}
  P(\ss, \lb|n) = \int P(\ss, \lb|\p,q,n)P(\p|n)P(q)\;\dd\p\,\dd q
\end{equation}
using uniform priors $P(q) = 1$, and
\begin{equation}
  P(\p|n) = \prod_lP(\p_l|d_l,n)P(d_l),
\end{equation}
with the uniform prior
\begin{equation}
  P(\p_l|d_l,n) = \prod_{\x}(D_l-1)!\delta{\textstyle\left(\sum_xp_{x,\x}^{l}-1\right)}.
\end{equation}
and
\begin{equation}
  P(d_l) = 2^{-D}
\end{equation}
being the prior for the alphabet $d_l$ of size $D_l$ inside segment $l$,
sampled uniformly from all possible subsets of the overall alphabet of
size $D$. Performing the above integral, we obtain
\begin{multline}\label{eq:joint_likelihood}
  P(\x, \lb|n) =  2^{-D(M+1)}\frac{M!(E-M)!}{(E+1)!}\times\\ \prod_l\prod_{\x}\frac{(D_l-1)!}{(k_{\x}^l+D_l-1)!}\prod_x a_{x,\x}^l!.
\end{multline}

Like with the previous stationary model, both the order and the
positions of the change points can be inferred from the joint posterior
distribution
\begin{equation}
  P(\lb,n|\x) = \frac{P(\x, \lb|n)P(n)}{P(\x)},
\end{equation}
in a manner that intrinsically prevents overfitting. This constitutes a
robust and elegant way of extracting this information from data, that
contrasts with non-Bayesian methods of detecting change points using
Markov chains that tend to be more
cumbersome~\cite{polansky_detecting_2007}, and is more versatile than
approaches that have a fixed Markov order~\cite{arnesen_bayesian_2016}.

The exact computation of the posterior of Eq.~\ref{eq:posterior_cp}
would require the marginalization of the above distribution for all
possible segments $\lb$, yielding the denominator $P(\x|n)$, which is
unfeasible for all but the smallest time series. However, it is not
necessary to compute this value if we sample $\lb$ from the posterior
using Monte Carlo. We do so by making move proposals $\lb\to\lb'$ with
probability $P(\lb'|\lb)$, and accepting it with probability $a$
according to the Metropolis-Hastings criterion~\cite{metropolis_equation_1953,hastings_monte_1970}
\begin{equation}
  a = \min\left(1,\frac{P(\lb'|\x,n)P(\lb|\lb')}{P(\lb|\x,n)P(\lb'|\lb)}\right),
\end{equation}
which does not require the computation of $P(\x|n)$ as it cancels out in
the ratio. If the move proposals are ergodic, i.e. they allow every
possible partition $\lb$ to be visited eventually, this algorithm will
asymptotically sample from the desired posterior. Here we use the
following move proposal scheme, choosing between one the following
actions with equal probability:
\begin{enumerate}
  \item We select a segment randomly and split it
        in a random point in the middle.
  \item We merge two adjacent segments.
  \item We move a randomly chosen boundary to a
        random position between the two enclosing ones.
\end{enumerate}
We perform this algorithm many times, starting from a single segment,
and waiting sufficiently long for equilibration --- determined by
observing if the likelihood value no longer changes significantly ---
and we choose the partition with the largest probability across
runs. For all datasets we investigated, we observed a fast convergence
of this algorithm, which typically shows very little variation between
runs.

Note that it is also possible to change the Markov order during the
algorithm, by proposing moves $n\to n'$, and using the
Metropolis-Hastings criterion to accept or reject them. However, we
found that Markov order typically settles very early in the algorithm,
and no longer changes during the remaining run, as it incurs a
macroscopic change in the likelihood. Since changing the Markov order is
an expensive operation of order $O(E)$, we have found it is best to
leave it fixed during the MCMC, and select it later according to the
likelihood value.

\begin{figure}[h]
  \includegraphics[width=\columnwidth]{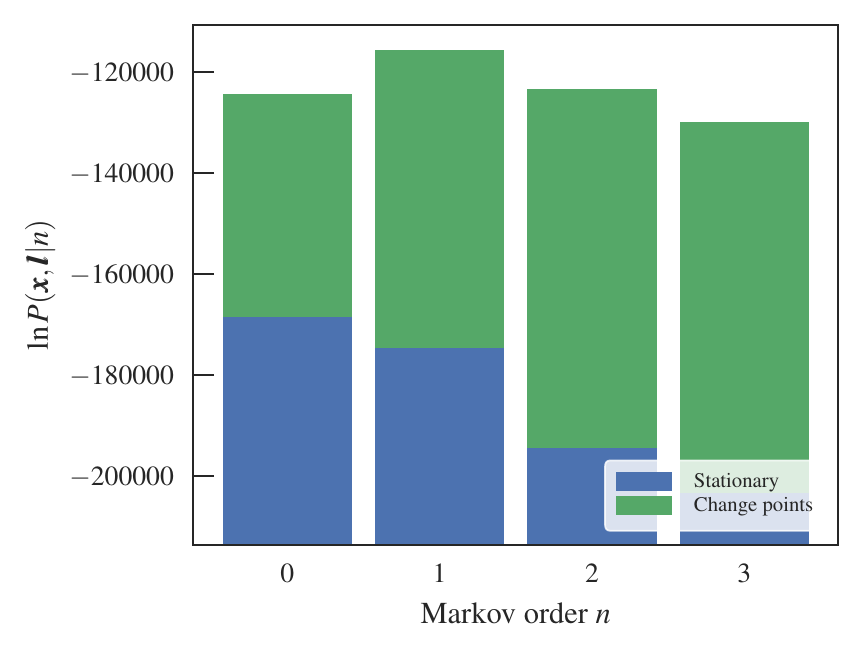} \caption{Integrated
  joint likelihood $P(\x,\lb|n)$ (Eq.~\ref{eq:joint_likelihood}) for a temporal network between students
  in a high school~\cite{fournet_contact_2014}, for the stationary
  (i.e. zero change points) and nonstationary models. For all values of
  $n$ the likelihoods are higher for the nonstationary model
    (yielding a posterior odds ratio $\Lambda>1$).\label{fig:comparison}}
\end{figure}

Once a fit is obtained, we can compare the above model with the
stationary one by computing the posterior odds ratio
\begin{equation}
  \Lambda = \frac{P(\lb,n|\x)}{P(\lb_0,n_0|\x)} = \frac{P(\x,\lb|n)}{P(\x,\lb_0|n_0)},
\end{equation}
where $\lb_0$ is the partition into a single interval (which is
equivalent to the stationary model). A value $\Lambda>1$
[i.e. $P(\x,\lb|n) > P(\x,\lb_0|n_0)$] indicates a larger evidence for
the nonstationary model.  As can be seen in Fig.~\ref{fig:comparison},
we observe indeed a larger evidence for the nonstationary model for all
Markov orders. In addition to this, using this general model we identify
$n=1$ as the most plausible Markov order, in contrast to the $n=0$
obtained with the stationary model. Therefore, identifying change points
allows us not only to uncover patterns at longer time scales, but the
separation into temporal segments enables the identification of
statistically significant patterns at short time scales as well, which
would otherwise remain obscured with the stationary model --- even
though it is designed to capture only these kinds of correlations.

\begin{figure}[h]
  \begin{overpic}[width=\columnwidth]{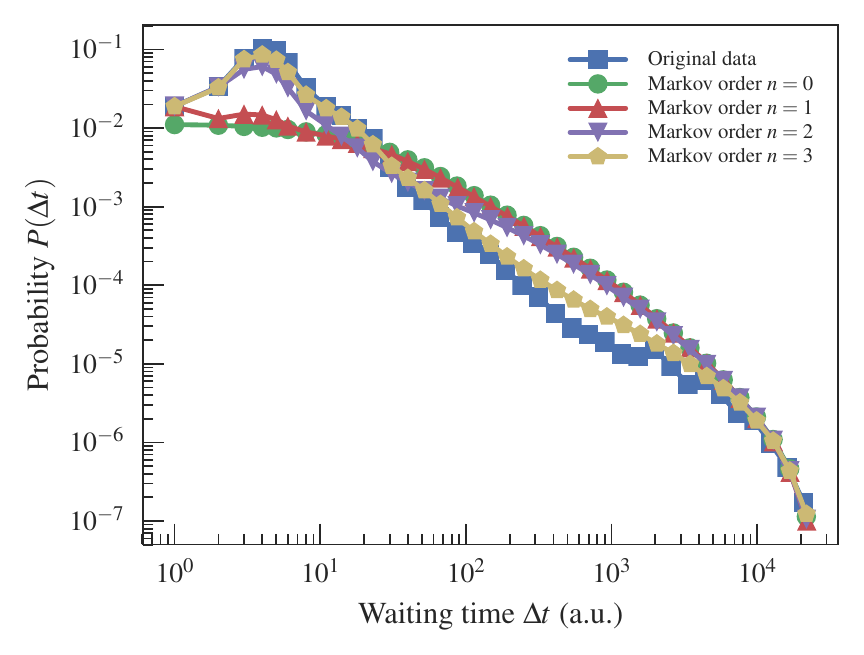}\put(0,5){(a)}\end{overpic}\\
  \begin{overpic}[width=\columnwidth]{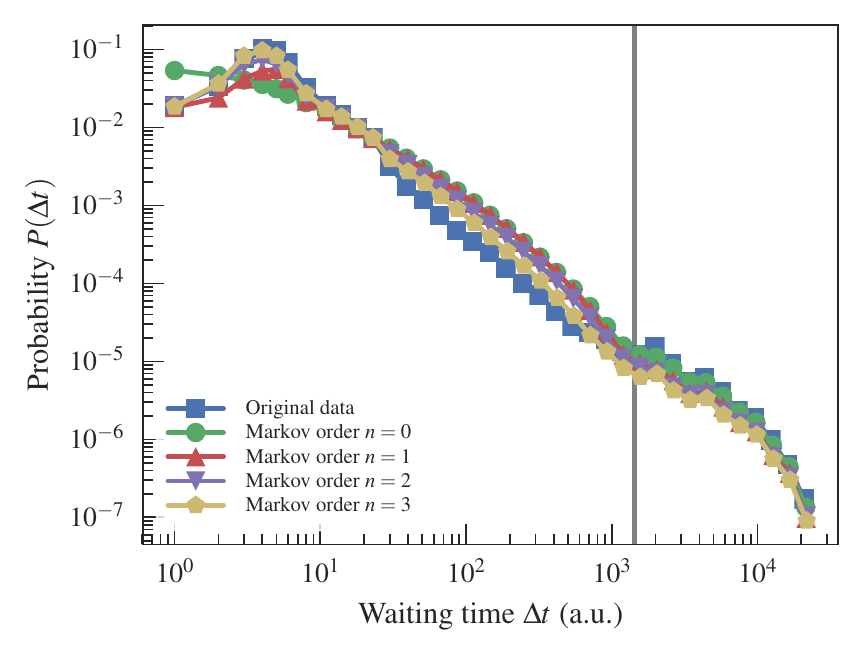}\put(0,5){(b)}\end{overpic}
  \caption{Distribution of waiting times $\Delta t$ between the same
  edge for the empirical dataset and fitted (a) stationary and (b)
  nonstationary models (a single instance of each), for a temporal
  network between students in a high
  school~\cite{fournet_contact_2014}. The vertical line shows the
  average length of inferred stationary segments between change
  points.\label{fig:waiting}}
\end{figure}

The improved quality of this model is also evident when we investigate
the epidemic dynamics, as shown in
Fig.~\ref{fig:markov-changepoints}. In order to obtain an estimate of
the number of infected based on the model, we generated different
sequences of edges using the fitted segments and transition
probabilities $\hat{p}_{x,\x}^{l} = a_{x,\x}^l / k_{\x}^l$ in each of the
segments estimated with Markov orders going from $0$ to $3$. We
simulated SIR and SIS processes on top of the networks generated and
averaged the number of infected over many instances. Looking at
Fig.~\ref{fig:markov-changepoints}, we see that the inferred positions
of the change-points tend to coincide with the abrupt changes in
infection rates, which show very good agreement between the empirical
and generated time-series. For higher Markov order, the agreement
improves, although the improvement seen for $n>1$ is probably due to
overfitting, given the results of Fig.~\ref{fig:comparison}. We note
also that the fact that $n=0$ provides the worse fit and agreement with
epidemic dynamics shows that it is not only the existence of change
points, but also the inferred Markov dynamics that contribute to the
quality of the model in reproducing the epidemic spreading.

In order to examine the link between the structure of the network and
the change points, we fitted a layered hierarchical degree-corrected
stochastic block
model~\cite{peixoto_inferring_2015,peixoto_nonparametric_2017} to the
data, considering each segment as a separate edge layer. From the figure
Fig.~\ref{fig:markov-changepoints}) we can see that the density of
connections between node groups vary in a substantial manner, suggesting
that change point marks an abrupt transition in the typical kind of
encounters between students --- representing breaks between classes,
meal time, etc (see Fig.~\ref{fig:markov-changepoints}). This yields an
insight as to why these changes in pattern may slow down or speed up an
epidemic spreading: if students are confined to their classrooms,
contagion across classrooms is inhibited, but as soon they are free to
move around the school grounds, so can the epidemic.

We explore further the match between data and model by measuring the
distribution of waiting times between temporal edges, i.e. the time
interval between the occurrence in the time series of the same edge in
the network, shown in Fig.~\ref{fig:waiting} for both Markov models. For
the empirical dataset, the waiting time distribution shows a
characteristic peak at short times, and a broad decay for longer ones.
For the stationary model, the distributions obtained with the fitted
models show significant discrepancy --- for both long and short times
--- except when the Markov order is increased to $n=3$, which, according
to our Bayesian analysis cannot be used as an explanation for the data,
as it represents an overfit. However, for the nonstationary model with
change points, we observe a fair agreement between data and model for
the most-likely model with $n=1$, across all time scales. The
nonstationary model also provides an explanation to the shape of the
distribution at longer times, which shows a separation of time scales
inside individual stationary segments, from larger ones across change
points (marked as vertical line in Fig.~\ref{fig:waiting}). In addition
to this, the fact that the $n=0$ model does not reproduce the short time
behavior of the distribution shows that the Markov property inside each
stationary segment is indeed a necessary ingredient of the model. The
model that best fits the data is able to reproduce with a quite good
degree of approximation the distribution of waiting times, across all
time scales. This point is in agreement with previous results
highlighting the importance of the heterogeneity of inter-event times
for dynamical processes~\cite{karsai2017bursty}, but here we see how two different time
scales are sufficient to reproduce a large fraction of the observed
behavior.

In appendix~\ref{app:datasets} we show that the same behavior is
obtained for a variety of different datasets.

\section{Discussion}\label{sec:conclusion}

In this work we presented a data-driven approach to model temporal
networks that is based on the simultaneous description of the network
dynamics in two time scales: 1. The occurrence of the edges according to
an arbitrary-order Markov chain, 2. The abrupt transition of the Markov
transition probabilities at specific change-points. We developed a
Bayesian framework that allows the inference of the change points and
Markov order from data in manner that prevents overfitting, and enables
the selection of competing models.

We have applied our approach to a variety of empirical proximity
networks, and we have evaluated the inferred models based on their
capacity to reproduce the epidemic spreading observed with the original
data. We have seen that the nonstationary model accurately reproduces
the highly-variable nature of the infection rate, with changes
correlating strongly with the inferred change points. Furthermore, we
showed that the inferred model also accurately reproduces the waiting
time statistics in the empirical data, both at small and large time
scales, neither of which are accurately captured if the different time
scales are analyzed in isolation.

We argue that, ultimately, the incorporation of such temporal
heterogeneity is indispensable for the evaluation of the speeding up or
slowing down of processes taking place on dynamic
networks~\cite{masuda_temporal_2013,scholtes_causality-driven_2014}, and
the development of mitigating strategies against
epidemics~\cite{gemmetto_mitigation_2014}.

Although our model successfully captures key properties of real dynamic
networks, it can still be made more realistic in a variety of ways. For
instance, it can be extended to continuous time via the incorporation of
waiting time distributions between events, as done in
Ref.~\cite{peixoto_modelling_2017}. Furthermore, it remains also to be
seen how the approach presented here can be extended to scenarios where
edges are allowed both to appear and disappear from the network, so that
its dynamics can no longer be represented simply by a sequence of
edges. And lastly, it would be desirable to provide a more direct
connection between the edge probabilities and change points with
large-scale network descriptors, such as community structure.

\begin{acknowledgments}
  The authors thank Ciro Cattuto, André Panisson and Anna Sapienza for useful
  conversations.
\end{acknowledgments}


\bibliography{bib,change_points}

\begin{thebibliography}{36}%
\makeatletter
\providecommand \@ifxundefined [1]{%
 \@ifx{#1\undefined}
}%
\providecommand \@ifnum [1]{%
 \ifnum #1\expandafter \@firstoftwo
 \else \expandafter \@secondoftwo
 \fi
}%
\providecommand \@ifx [1]{%
 \ifx #1\expandafter \@firstoftwo
 \else \expandafter \@secondoftwo
 \fi
}%
\providecommand \natexlab [1]{#1}%
\providecommand \enquote  [1]{``#1''}%
\providecommand \bibnamefont  [1]{#1}%
\providecommand \bibfnamefont [1]{#1}%
\providecommand \citenamefont [1]{#1}%
\providecommand \href@noop [0]{\@secondoftwo}%
\providecommand \href [0]{\begingroup \@sanitize@url \@href}%
\providecommand \@href[1]{\@@startlink{#1}\@@href}%
\providecommand \@@href[1]{\endgroup#1\@@endlink}%
\providecommand \@sanitize@url [0]{\catcode `\\12\catcode `\$12\catcode
  `\&12\catcode `\#12\catcode `\^12\catcode `\_12\catcode `\%12\relax}%
\providecommand \@@startlink[1]{}%
\providecommand \@@endlink[0]{}%
\providecommand \url  [0]{\begingroup\@sanitize@url \@url }%
\providecommand \@url [1]{\endgroup\@href {#1}{\urlprefix }}%
\providecommand \urlprefix  [0]{URL }%
\providecommand \Eprint [0]{\href }%
\providecommand \doibase [0]{http://dx.doi.org/}%
\providecommand \selectlanguage [0]{\@gobble}%
\providecommand \bibinfo  [0]{\@secondoftwo}%
\providecommand \bibfield  [0]{\@secondoftwo}%
\providecommand \translation [1]{[#1]}%
\providecommand \BibitemOpen [0]{}%
\providecommand \bibitemStop [0]{}%
\providecommand \bibitemNoStop [0]{.\EOS\space}%
\providecommand \EOS [0]{\spacefactor3000\relax}%
\providecommand \BibitemShut  [1]{\csname bibitem#1\endcsname}%
\let\auto@bib@innerbib\@empty
\bibitem [{\citenamefont {Holme}\ and\ \citenamefont
  {Saramäki}(2012)}]{holme_temporal_2012}%
  \BibitemOpen
  \bibfield  {author} {\bibinfo {author} {\bibfnamefont {Petter}\ \bibnamefont
  {Holme}}\ and\ \bibinfo {author} {\bibfnamefont {Jari}\ \bibnamefont
  {Saramäki}},\ }\bibfield  {title} {\enquote {\bibinfo {title} {Temporal
  networks},}\ }\href {\doibase 10.1016/j.physrep.2012.03.001} {\bibfield
  {journal} {\bibinfo  {journal} {Physics Reports}\ }\textbf {\bibinfo {volume}
  {519}},\ \bibinfo {pages} {97--125} (\bibinfo {year} {2012})}\BibitemShut
  {NoStop}%
\bibitem [{\citenamefont {Holme}(2015)}]{holme_modern_2015}%
  \BibitemOpen
  \bibfield  {author} {\bibinfo {author} {\bibfnamefont {Petter}\ \bibnamefont
  {Holme}},\ }\bibfield  {title} {{\selectlanguage {english}\enquote {\bibinfo
  {title} {Modern temporal network theory: a colloquium},}\ }}\href {\doibase
  10.1140/epjb/e2015-60657-4} {\bibfield  {journal} {\bibinfo  {journal} {The
  European Physical Journal B}\ }\textbf {\bibinfo {volume} {88}},\ \bibinfo
  {pages} {234} (\bibinfo {year} {2015})}\BibitemShut {NoStop}%
\bibitem [{\citenamefont {Ho}\ \emph {et~al.}(2011)\citenamefont {Ho},
  \citenamefont {Song},\ and\ \citenamefont {Xing}}]{ho2011evolving}%
  \BibitemOpen
  \bibfield  {author} {\bibinfo {author} {\bibfnamefont {Qirong}\ \bibnamefont
  {Ho}}, \bibinfo {author} {\bibfnamefont {Le}~\bibnamefont {Song}}, \ and\
  \bibinfo {author} {\bibfnamefont {Eric~P}\ \bibnamefont {Xing}},\ }\bibfield
  {title} {\enquote {\bibinfo {title} {Evolving cluster mixed-membership
  blockmodel for time-varying networks},}\ }\href@noop {} {\bibfield  {journal}
  {\bibinfo  {journal} {Journal of Machine Learning Research : Workshop and
  Conference Proceedings}\ ,\ \bibinfo {pages} {342--350}} (\bibinfo {year}
  {2011})}\BibitemShut {NoStop}%
\bibitem [{\citenamefont {Perra}\ \emph {et~al.}(2012)\citenamefont {Perra},
  \citenamefont {Gon{\c{c}}alves}, \citenamefont {Pastor-Satorras},\ and\
  \citenamefont {Vespignani}}]{perra2012activity}%
  \BibitemOpen
  \bibfield  {author} {\bibinfo {author} {\bibfnamefont {Nicola}\ \bibnamefont
  {Perra}}, \bibinfo {author} {\bibfnamefont {Bruno}\ \bibnamefont
  {Gon{\c{c}}alves}}, \bibinfo {author} {\bibfnamefont {Romualdo}\ \bibnamefont
  {Pastor-Satorras}}, \ and\ \bibinfo {author} {\bibfnamefont {Alessandro}\
  \bibnamefont {Vespignani}},\ }\bibfield  {title} {\enquote {\bibinfo {title}
  {Activity driven modeling of time varying networks},}\ }\href@noop {}
  {\bibfield  {journal} {\bibinfo  {journal} {Scientific reports}\ }\textbf
  {\bibinfo {volume} {2}} (\bibinfo {year} {2012})}\BibitemShut {NoStop}%
\bibitem [{\citenamefont {Rocha}\ \emph {et~al.}(2011)\citenamefont {Rocha},
  \citenamefont {Liljeros},\ and\ \citenamefont
  {Holme}}]{rocha_simulated_2011}%
  \BibitemOpen
  \bibfield  {author} {\bibinfo {author} {\bibfnamefont {Luis E.~C.}\
  \bibnamefont {Rocha}}, \bibinfo {author} {\bibfnamefont {Fredrik}\
  \bibnamefont {Liljeros}}, \ and\ \bibinfo {author} {\bibfnamefont {Petter}\
  \bibnamefont {Holme}},\ }\bibfield  {title} {\enquote {\bibinfo {title}
  {Simulated {Epidemics} in an {Empirical} {Spatiotemporal} {Network} of 50,185
  {Sexual} {Contacts}},}\ }\href {\doibase 10.1371/journal.pcbi.1001109}
  {\bibfield  {journal} {\bibinfo  {journal} {PLOS Computational Biology}\
  }\textbf {\bibinfo {volume} {7}},\ \bibinfo {pages} {e1001109} (\bibinfo
  {year} {2011})}\BibitemShut {NoStop}%
\bibitem [{\citenamefont {Valdano}\ \emph {et~al.}(2015)\citenamefont
  {Valdano}, \citenamefont {Ferreri}, \citenamefont {Poletto},\ and\
  \citenamefont {Colizza}}]{valdano_analytical_2015}%
  \BibitemOpen
  \bibfield  {author} {\bibinfo {author} {\bibfnamefont {Eugenio}\ \bibnamefont
  {Valdano}}, \bibinfo {author} {\bibfnamefont {Luca}\ \bibnamefont {Ferreri}},
  \bibinfo {author} {\bibfnamefont {Chiara}\ \bibnamefont {Poletto}}, \ and\
  \bibinfo {author} {\bibfnamefont {Vittoria}\ \bibnamefont {Colizza}},\
  }\bibfield  {title} {\enquote {\bibinfo {title} {Analytical {Computation} of
  the {Epidemic} {Threshold} on {Temporal} {Networks}},}\ }\href {\doibase
  10.1103/PhysRevX.5.021005} {\bibfield  {journal} {\bibinfo  {journal}
  {Physical Review X}\ }\textbf {\bibinfo {volume} {5}},\ \bibinfo {pages}
  {021005} (\bibinfo {year} {2015})}\BibitemShut {NoStop}%
\bibitem [{\citenamefont {Génois}\ \emph {et~al.}(2015)\citenamefont
  {Génois}, \citenamefont {Vestergaard}, \citenamefont {Cattuto},\ and\
  \citenamefont {Barrat}}]{genois_compensating_2015}%
  \BibitemOpen
  \bibfield  {author} {\bibinfo {author} {\bibfnamefont {Mathieu}\ \bibnamefont
  {Génois}}, \bibinfo {author} {\bibfnamefont {Christian~L.}\ \bibnamefont
  {Vestergaard}}, \bibinfo {author} {\bibfnamefont {Ciro}\ \bibnamefont
  {Cattuto}}, \ and\ \bibinfo {author} {\bibfnamefont {Alain}\ \bibnamefont
  {Barrat}},\ }\bibfield  {title} {\enquote {\bibinfo {title} {Compensating for
  population sampling in simulations of epidemic spread on temporal contact
  networks},}\ }\href {\doibase 10.1038/ncomms9860} {\bibfield  {journal}
  {\bibinfo  {journal} {Nature Communications}\ }\textbf {\bibinfo {volume}
  {6}} (\bibinfo {year} {2015}),\ 10.1038/ncomms9860}\BibitemShut {NoStop}%
\bibitem [{\citenamefont {Ren}\ and\ \citenamefont
  {Wang}(2014)}]{ren2014epidemic}%
  \BibitemOpen
  \bibfield  {author} {\bibinfo {author} {\bibfnamefont {Guangming}\
  \bibnamefont {Ren}}\ and\ \bibinfo {author} {\bibfnamefont {Xingyuan}\
  \bibnamefont {Wang}},\ }\bibfield  {title} {\enquote {\bibinfo {title}
  {Epidemic spreading in time-varying community networks},}\ }\href@noop {}
  {\bibfield  {journal} {\bibinfo  {journal} {Chaos: An Interdisciplinary
  Journal of Nonlinear Science}\ }\textbf {\bibinfo {volume} {24}},\ \bibinfo
  {pages} {023116} (\bibinfo {year} {2014})}\BibitemShut {NoStop}%
\bibitem [{\citenamefont {Karsai}\ \emph {et~al.}(2011)\citenamefont {Karsai},
  \citenamefont {Kivel{\"a}}, \citenamefont {Pan}, \citenamefont {Kaski},
  \citenamefont {Kert{\'e}sz}, \citenamefont {Barab{\'a}si},\ and\
  \citenamefont {Saram{\"a}ki}}]{karsai2011small}%
  \BibitemOpen
  \bibfield  {author} {\bibinfo {author} {\bibfnamefont {M{\'a}rton}\
  \bibnamefont {Karsai}}, \bibinfo {author} {\bibfnamefont {Mikko}\
  \bibnamefont {Kivel{\"a}}}, \bibinfo {author} {\bibfnamefont {Raj~Kumar}\
  \bibnamefont {Pan}}, \bibinfo {author} {\bibfnamefont {Kimmo}\ \bibnamefont
  {Kaski}}, \bibinfo {author} {\bibfnamefont {J{\'a}nos}\ \bibnamefont
  {Kert{\'e}sz}}, \bibinfo {author} {\bibfnamefont {A-L}\ \bibnamefont
  {Barab{\'a}si}}, \ and\ \bibinfo {author} {\bibfnamefont {Jari}\ \bibnamefont
  {Saram{\"a}ki}},\ }\bibfield  {title} {\enquote {\bibinfo {title} {Small but
  slow world: How network topology and burstiness slow down spreading},}\
  }\href@noop {} {\bibfield  {journal} {\bibinfo  {journal} {Physical Review
  E}\ }\textbf {\bibinfo {volume} {83}},\ \bibinfo {pages} {025102} (\bibinfo
  {year} {2011})}\BibitemShut {NoStop}%
\bibitem [{\citenamefont {Gauvin}\ \emph {et~al.}(2013)\citenamefont {Gauvin},
  \citenamefont {Panisson}, \citenamefont {Cattuto},\ and\ \citenamefont
  {Barrat}}]{gauvin2013activity}%
  \BibitemOpen
  \bibfield  {author} {\bibinfo {author} {\bibfnamefont {Laetitia}\
  \bibnamefont {Gauvin}}, \bibinfo {author} {\bibfnamefont {Andr{\'e}}\
  \bibnamefont {Panisson}}, \bibinfo {author} {\bibfnamefont {Ciro}\
  \bibnamefont {Cattuto}}, \ and\ \bibinfo {author} {\bibfnamefont {Alain}\
  \bibnamefont {Barrat}},\ }\bibfield  {title} {\enquote {\bibinfo {title}
  {Activity clocks: spreading dynamics on temporal networks of human
  contact},}\ }\href@noop {} {\bibfield  {journal} {\bibinfo  {journal}
  {Scientific reports}\ }\textbf {\bibinfo {volume} {3}} (\bibinfo {year}
  {2013})}\BibitemShut {NoStop}%
\bibitem [{\citenamefont {Vestergaard}\ \emph {et~al.}(2014)\citenamefont
  {Vestergaard}, \citenamefont {G{\'e}nois},\ and\ \citenamefont
  {Barrat}}]{vestergaard2014memory}%
  \BibitemOpen
  \bibfield  {author} {\bibinfo {author} {\bibfnamefont {Christian~L}\
  \bibnamefont {Vestergaard}}, \bibinfo {author} {\bibfnamefont {Mathieu}\
  \bibnamefont {G{\'e}nois}}, \ and\ \bibinfo {author} {\bibfnamefont {Alain}\
  \bibnamefont {Barrat}},\ }\bibfield  {title} {\enquote {\bibinfo {title} {How
  memory generates heterogeneous dynamics in temporal networks},}\ }\href@noop
  {} {\bibfield  {journal} {\bibinfo  {journal} {Physical Review E}\ }\textbf
  {\bibinfo {volume} {90}},\ \bibinfo {pages} {042805} (\bibinfo {year}
  {2014})}\BibitemShut {NoStop}%
\bibitem [{\citenamefont {Scholtes}\ \emph {et~al.}(2014)\citenamefont
  {Scholtes}, \citenamefont {Wider}, \citenamefont {Pfitzner}, \citenamefont
  {Garas}, \citenamefont {Tessone},\ and\ \citenamefont
  {Schweitzer}}]{scholtes_causality-driven_2014}%
  \BibitemOpen
  \bibfield  {author} {\bibinfo {author} {\bibfnamefont {Ingo}\ \bibnamefont
  {Scholtes}}, \bibinfo {author} {\bibfnamefont {Nicolas}\ \bibnamefont
  {Wider}}, \bibinfo {author} {\bibfnamefont {René}\ \bibnamefont {Pfitzner}},
  \bibinfo {author} {\bibfnamefont {Antonios}\ \bibnamefont {Garas}}, \bibinfo
  {author} {\bibfnamefont {Claudio~J.}\ \bibnamefont {Tessone}}, \ and\
  \bibinfo {author} {\bibfnamefont {Frank}\ \bibnamefont {Schweitzer}},\
  }\bibfield  {title} {{\selectlanguage {english}\enquote {\bibinfo {title}
  {Causality-driven slow-down and speed-up of diffusion in non-{Markovian}
  temporal networks},}\ }}\href {\doibase 10.1038/ncomms6024} {\bibfield
  {journal} {\bibinfo  {journal} {Nature Communications}\ }\textbf {\bibinfo
  {volume} {5}} (\bibinfo {year} {2014}),\ 10.1038/ncomms6024}\BibitemShut
  {NoStop}%
\bibitem [{\citenamefont {Peixoto}\ and\ \citenamefont
  {Rosvall}(2017)}]{peixoto_modelling_2017}%
  \BibitemOpen
  \bibfield  {author} {\bibinfo {author} {\bibfnamefont {Tiago~P.}\
  \bibnamefont {Peixoto}}\ and\ \bibinfo {author} {\bibfnamefont {Martin}\
  \bibnamefont {Rosvall}},\ }\bibfield  {title} {{\selectlanguage
  {english}\enquote {\bibinfo {title} {Modelling sequences and temporal
  networks with dynamic community structures},}\ }}\href {\doibase
  10.1038/s41467-017-00148-9} {\bibfield  {journal} {\bibinfo  {journal}
  {Nature Communications}\ }\textbf {\bibinfo {volume} {8}},\ \bibinfo {pages}
  {582} (\bibinfo {year} {2017})}\BibitemShut {NoStop}%
\bibitem [{\citenamefont {Xu}\ and\ \citenamefont
  {Iii}(2013)}]{xu_dynamic_2013}%
  \BibitemOpen
  \bibfield  {author} {\bibinfo {author} {\bibfnamefont {Kevin~S.}\
  \bibnamefont {Xu}}\ and\ \bibinfo {author} {\bibfnamefont {Alfred O.~Hero}\
  \bibnamefont {Iii}},\ }\bibfield  {title} {\enquote {\bibinfo {title}
  {Dynamic {Stochastic} {Blockmodels}: {Statistical} {Models} for
  {Time}-{Evolving} {Networks}},}\ }in\ \href
  {http://link.springer.com/chapter/10.1007/978-3-642-37210-0_22} {\emph
  {\bibinfo {booktitle} {Social {Computing}, {Behavioral}-{Cultural} {Modeling}
  and {Prediction}}}},\ \bibinfo {series and number} {\bibinfo {series}
  {Lecture {Notes} in {Computer} {Science}}\ No.\ \bibinfo {number} {7812}},\
  \bibinfo {editor} {edited by\ \bibinfo {editor} {\bibfnamefont {Ariel~M.}\
  \bibnamefont {Greenberg}}, \bibinfo {editor} {\bibfnamefont {William~G.}\
  \bibnamefont {Kennedy}}, \ and\ \bibinfo {editor} {\bibfnamefont {Nathan~D.}\
  \bibnamefont {Bos}}}\ (\bibinfo  {publisher} {Springer Berlin Heidelberg},\
  \bibinfo {year} {2013})\ pp.\ \bibinfo {pages} {201--210}\BibitemShut
  {NoStop}%
\bibitem [{\citenamefont {Gauvin}\ \emph {et~al.}(2014)\citenamefont {Gauvin},
  \citenamefont {Panisson},\ and\ \citenamefont
  {Cattuto}}]{gauvin_detecting_2014}%
  \BibitemOpen
  \bibfield  {author} {\bibinfo {author} {\bibfnamefont {Laetitia}\
  \bibnamefont {Gauvin}}, \bibinfo {author} {\bibfnamefont {André}\
  \bibnamefont {Panisson}}, \ and\ \bibinfo {author} {\bibfnamefont {Ciro}\
  \bibnamefont {Cattuto}},\ }\bibfield  {title} {\enquote {\bibinfo {title}
  {Detecting the {Community} {Structure} and {Activity} {Patterns} of
  {Temporal} {Networks}: {A} {Non}-{Negative} {Tensor} {Factorization}
  {Approach}},}\ }\href {\doibase 10.1371/journal.pone.0086028} {\bibfield
  {journal} {\bibinfo  {journal} {PLoS ONE}\ }\textbf {\bibinfo {volume} {9}},\
  \bibinfo {pages} {e86028} (\bibinfo {year} {2014})}\BibitemShut {NoStop}%
\bibitem [{\citenamefont {Peixoto}(2015)}]{peixoto_inferring_2015}%
  \BibitemOpen
  \bibfield  {author} {\bibinfo {author} {\bibfnamefont {Tiago~P.}\
  \bibnamefont {Peixoto}},\ }\bibfield  {title} {\enquote {\bibinfo {title}
  {Inferring the mesoscale structure of layered, edge-valued, and time-varying
  networks},}\ }\href {\doibase 10.1103/PhysRevE.92.042807} {\bibfield
  {journal} {\bibinfo  {journal} {Physical Review E}\ }\textbf {\bibinfo
  {volume} {92}},\ \bibinfo {pages} {042807} (\bibinfo {year}
  {2015})}\BibitemShut {NoStop}%
\bibitem [{\citenamefont {Stanley}\ \emph {et~al.}(2016)\citenamefont
  {Stanley}, \citenamefont {Shai}, \citenamefont {Taylor},\ and\ \citenamefont
  {Mucha}}]{stanley_clustering_2016}%
  \BibitemOpen
  \bibfield  {author} {\bibinfo {author} {\bibfnamefont {N.}~\bibnamefont
  {Stanley}}, \bibinfo {author} {\bibfnamefont {S.}~\bibnamefont {Shai}},
  \bibinfo {author} {\bibfnamefont {D.}~\bibnamefont {Taylor}}, \ and\ \bibinfo
  {author} {\bibfnamefont {P.~J.}\ \bibnamefont {Mucha}},\ }\bibfield  {title}
  {\enquote {\bibinfo {title} {Clustering {Network} {Layers} with the {Strata}
  {Multilayer} {Stochastic} {Block} {Model}},}\ }\href {\doibase
  10.1109/TNSE.2016.2537545} {\bibfield  {journal} {\bibinfo  {journal} {IEEE
  Transactions on Network Science and Engineering}\ }\textbf {\bibinfo {volume}
  {3}},\ \bibinfo {pages} {95--105} (\bibinfo {year} {2016})}\BibitemShut
  {NoStop}%
\bibitem [{\citenamefont {Ghasemian}\ \emph {et~al.}(2016)\citenamefont
  {Ghasemian}, \citenamefont {Zhang}, \citenamefont {Clauset}, \citenamefont
  {Moore},\ and\ \citenamefont {Peel}}]{ghasemian_detectability_2016}%
  \BibitemOpen
  \bibfield  {author} {\bibinfo {author} {\bibfnamefont {Amir}\ \bibnamefont
  {Ghasemian}}, \bibinfo {author} {\bibfnamefont {Pan}\ \bibnamefont {Zhang}},
  \bibinfo {author} {\bibfnamefont {Aaron}\ \bibnamefont {Clauset}}, \bibinfo
  {author} {\bibfnamefont {Cristopher}\ \bibnamefont {Moore}}, \ and\ \bibinfo
  {author} {\bibfnamefont {Leto}\ \bibnamefont {Peel}},\ }\bibfield  {title}
  {\enquote {\bibinfo {title} {Detectability {Thresholds} and {Optimal}
  {Algorithms} for {Community} {Structure} in {Dynamic} {Networks}},}\ }\href
  {\doibase 10.1103/PhysRevX.6.031005} {\bibfield  {journal} {\bibinfo
  {journal} {Physical Review X}\ }\textbf {\bibinfo {volume} {6}},\ \bibinfo
  {pages} {031005} (\bibinfo {year} {2016})}\BibitemShut {NoStop}%
\bibitem [{\citenamefont {Zhang}\ \emph {et~al.}(2017)\citenamefont {Zhang},
  \citenamefont {Moore},\ and\ \citenamefont {Newman}}]{zhang_random_2017}%
  \BibitemOpen
  \bibfield  {author} {\bibinfo {author} {\bibfnamefont {Xiao}\ \bibnamefont
  {Zhang}}, \bibinfo {author} {\bibfnamefont {Cristopher}\ \bibnamefont
  {Moore}}, \ and\ \bibinfo {author} {\bibfnamefont {Mark E.~J.}\ \bibnamefont
  {Newman}},\ }\bibfield  {title} {{\selectlanguage {english}\enquote {\bibinfo
  {title} {Random graph models for dynamic networks},}\ }}\href {\doibase
  10.1140/epjb/e2017-80122-8} {\bibfield  {journal} {\bibinfo  {journal} {The
  European Physical Journal B}\ }\textbf {\bibinfo {volume} {90}},\ \bibinfo
  {pages} {200} (\bibinfo {year} {2017})}\BibitemShut {NoStop}%
\bibitem [{\citenamefont {Peel}\ and\ \citenamefont
  {Clauset}(2015)}]{peel_detecting_2015}%
  \BibitemOpen
  \bibfield  {author} {\bibinfo {author} {\bibfnamefont {Leto}\ \bibnamefont
  {Peel}}\ and\ \bibinfo {author} {\bibfnamefont {Aaron}\ \bibnamefont
  {Clauset}},\ }\bibfield  {title} {{\selectlanguage {english}\enquote
  {\bibinfo {title} {Detecting {Change} {Points} in the {Large}-{Scale}
  {Structure} of {Evolving} {Networks}},}\ }}in\ \href
  {https://www.aaai.org/ocs/index.php/AAAI/AAAI15/paper/view/9485}
  {{\selectlanguage {english}\emph {\bibinfo {booktitle} {Twenty-{Ninth} {AAAI}
  {Conference} on {Artificial} {Intelligence}}}}}\ (\bibinfo {year}
  {2015})\BibitemShut {NoStop}%
\bibitem [{\citenamefont {De~Ridder}\ \emph {et~al.}(2016)\citenamefont
  {De~Ridder}, \citenamefont {Vandermarliere},\ and\ \citenamefont
  {Ryckebusch}}]{de2016detection}%
  \BibitemOpen
  \bibfield  {author} {\bibinfo {author} {\bibfnamefont {Simon}\ \bibnamefont
  {De~Ridder}}, \bibinfo {author} {\bibfnamefont {Benjamin}\ \bibnamefont
  {Vandermarliere}}, \ and\ \bibinfo {author} {\bibfnamefont {Jan}\
  \bibnamefont {Ryckebusch}},\ }\bibfield  {title} {\enquote {\bibinfo {title}
  {Detection and localization of change points in temporal networks with the
  aid of stochastic block models},}\ }\href@noop {} {\bibfield  {journal}
  {\bibinfo  {journal} {Journal of Statistical Mechanics: Theory and
  Experiment}\ }\textbf {\bibinfo {volume} {2016}},\ \bibinfo {pages} {113302}
  (\bibinfo {year} {2016})}\BibitemShut {NoStop}%
\bibitem [{\citenamefont {Corneli}\ \emph {et~al.}(2017)\citenamefont
  {Corneli}, \citenamefont {Latouche},\ and\ \citenamefont
  {Rossi}}]{cornelimultiple}%
  \BibitemOpen
  \bibfield  {author} {\bibinfo {author} {\bibfnamefont {Marco}\ \bibnamefont
  {Corneli}}, \bibinfo {author} {\bibfnamefont {Pierre}\ \bibnamefont
  {Latouche}}, \ and\ \bibinfo {author} {\bibfnamefont {Fabrice}\ \bibnamefont
  {Rossi}},\ }\bibfield  {title} {\enquote {\bibinfo {title} {Multiple change
  points detection and clustering in dynamic network},}\ }\href@noop {} {\
  (\bibinfo {year} {2017})}\BibitemShut {NoStop}%
\bibitem [{\citenamefont {Toroczkai}\ and\ \citenamefont
  {Guclu}(2007)}]{toroczkai_proximity_2007}%
  \BibitemOpen
  \bibfield  {author} {\bibinfo {author} {\bibfnamefont {Zoltán}\ \bibnamefont
  {Toroczkai}}\ and\ \bibinfo {author} {\bibfnamefont {Hasan}\ \bibnamefont
  {Guclu}},\ }\bibfield  {title} {\enquote {\bibinfo {title} {Proximity
  networks and epidemics},}\ }\href {\doibase 10.1016/j.physa.2006.11.088}
  {\bibfield  {journal} {\bibinfo  {journal} {Physica A: Statistical Mechanics
  and its Applications}\ }\bibinfo {series} {Social network analysis:
  {Measuring} tools, structures and dynamics},\ \textbf {\bibinfo {volume}
  {378}},\ \bibinfo {pages} {68--75} (\bibinfo {year} {2007})}\BibitemShut
  {NoStop}%
\bibitem [{\citenamefont {Stehlé}\ \emph {et~al.}(2011)\citenamefont
  {Stehlé}, \citenamefont {Voirin}, \citenamefont {Barrat}, \citenamefont
  {Cattuto}, \citenamefont {Isella}, \citenamefont {Pinton}, \citenamefont
  {Quaggiotto}, \citenamefont {Broeck}, \citenamefont {Régis}, \citenamefont
  {Lina},\ and\ \citenamefont {Vanhems}}]{stehle_high-resolution_2011}%
  \BibitemOpen
  \bibfield  {author} {\bibinfo {author} {\bibfnamefont {Juliette}\
  \bibnamefont {Stehlé}}, \bibinfo {author} {\bibfnamefont {Nicolas}\
  \bibnamefont {Voirin}}, \bibinfo {author} {\bibfnamefont {Alain}\
  \bibnamefont {Barrat}}, \bibinfo {author} {\bibfnamefont {Ciro}\ \bibnamefont
  {Cattuto}}, \bibinfo {author} {\bibfnamefont {Lorenzo}\ \bibnamefont
  {Isella}}, \bibinfo {author} {\bibfnamefont {Jean-François}\ \bibnamefont
  {Pinton}}, \bibinfo {author} {\bibfnamefont {Marco}\ \bibnamefont
  {Quaggiotto}}, \bibinfo {author} {\bibfnamefont {Wouter Van~den}\
  \bibnamefont {Broeck}}, \bibinfo {author} {\bibfnamefont {Corinne}\
  \bibnamefont {Régis}}, \bibinfo {author} {\bibfnamefont {Bruno}\
  \bibnamefont {Lina}}, \ and\ \bibinfo {author} {\bibfnamefont {Philippe}\
  \bibnamefont {Vanhems}},\ }\bibfield  {title} {\enquote {\bibinfo {title}
  {High-{Resolution} {Measurements} of {Face}-to-{Face} {Contact} {Patterns} in
  a {Primary} {School}},}\ }\href {\doibase 10.1371/journal.pone.0023176}
  {\bibfield  {journal} {\bibinfo  {journal} {PLOS ONE}\ }\textbf {\bibinfo
  {volume} {6}},\ \bibinfo {pages} {e23176} (\bibinfo {year}
  {2011})}\BibitemShut {NoStop}%
\bibitem [{\citenamefont {Vanhems}\ \emph {et~al.}(2013)\citenamefont
  {Vanhems}, \citenamefont {Barrat}, \citenamefont {Cattuto}, \citenamefont
  {Pinton}, \citenamefont {Khanafer}, \citenamefont {Régis}, \citenamefont
  {Kim}, \citenamefont {Comte},\ and\ \citenamefont
  {Voirin}}]{vanhems_estimating_2013}%
  \BibitemOpen
  \bibfield  {author} {\bibinfo {author} {\bibfnamefont {Philippe}\
  \bibnamefont {Vanhems}}, \bibinfo {author} {\bibfnamefont {Alain}\
  \bibnamefont {Barrat}}, \bibinfo {author} {\bibfnamefont {Ciro}\ \bibnamefont
  {Cattuto}}, \bibinfo {author} {\bibfnamefont {Jean-François}\ \bibnamefont
  {Pinton}}, \bibinfo {author} {\bibfnamefont {Nagham}\ \bibnamefont
  {Khanafer}}, \bibinfo {author} {\bibfnamefont {Corinne}\ \bibnamefont
  {Régis}}, \bibinfo {author} {\bibfnamefont {Byeul-a}\ \bibnamefont {Kim}},
  \bibinfo {author} {\bibfnamefont {Brigitte}\ \bibnamefont {Comte}}, \ and\
  \bibinfo {author} {\bibfnamefont {Nicolas}\ \bibnamefont {Voirin}},\
  }\bibfield  {title} {\enquote {\bibinfo {title} {Estimating {Potential}
  {Infection} {Transmission} {Routes} in {Hospital} {Wards} {Using} {Wearable}
  {Proximity} {Sensors}},}\ }\href {\doibase 10.1371/journal.pone.0073970}
  {\bibfield  {journal} {\bibinfo  {journal} {PLoS ONE}\ }\textbf {\bibinfo
  {volume} {8}},\ \bibinfo {pages} {e73970} (\bibinfo {year}
  {2013})}\BibitemShut {NoStop}%
\bibitem [{\citenamefont {Mastrandrea}\ \emph {et~al.}(2015)\citenamefont
  {Mastrandrea}, \citenamefont {Fournet},\ and\ \citenamefont
  {Barrat}}]{mastrandrea_contact_2015}%
  \BibitemOpen
  \bibfield  {author} {\bibinfo {author} {\bibfnamefont {Rossana}\ \bibnamefont
  {Mastrandrea}}, \bibinfo {author} {\bibfnamefont {Julie}\ \bibnamefont
  {Fournet}}, \ and\ \bibinfo {author} {\bibfnamefont {Alain}\ \bibnamefont
  {Barrat}},\ }\bibfield  {title} {\enquote {\bibinfo {title} {Contact
  {Patterns} in a {High} {School}: {A} {Comparison} between {Data} {Collected}
  {Using} {Wearable} {Sensors}, {Contact} {Diaries} and {Friendship}
  {Surveys}},}\ }\href {\doibase 10.1371/journal.pone.0136497} {\bibfield
  {journal} {\bibinfo  {journal} {PLoS ONE}\ }\textbf {\bibinfo {volume}
  {10}},\ \bibinfo {pages} {e0136497} (\bibinfo {year} {2015})}\BibitemShut
  {NoStop}%
\bibitem [{\citenamefont {Gemmetto}\ \emph {et~al.}(2014)\citenamefont
  {Gemmetto}, \citenamefont {Barrat},\ and\ \citenamefont
  {Cattuto}}]{gemmetto_mitigation_2014}%
  \BibitemOpen
  \bibfield  {author} {\bibinfo {author} {\bibfnamefont {Valerio}\ \bibnamefont
  {Gemmetto}}, \bibinfo {author} {\bibfnamefont {Alain}\ \bibnamefont
  {Barrat}}, \ and\ \bibinfo {author} {\bibfnamefont {Ciro}\ \bibnamefont
  {Cattuto}},\ }\bibfield  {title} {{\selectlanguage {english}\enquote
  {\bibinfo {title} {Mitigation of infectious disease at school: targeted class
  closure vs school closure},}\ }}\href {\doibase 10.1186/s12879-014-0695-9}
  {\bibfield  {journal} {\bibinfo  {journal} {BMC Infectious Diseases}\
  }\textbf {\bibinfo {volume} {14}},\ \bibinfo {pages} {695} (\bibinfo {year}
  {2014})}\BibitemShut {NoStop}%
\bibitem [{\citenamefont {Fournet}\ and\ \citenamefont
  {Barrat}(2014)}]{fournet_contact_2014}%
  \BibitemOpen
  \bibfield  {author} {\bibinfo {author} {\bibfnamefont {Julie}\ \bibnamefont
  {Fournet}}\ and\ \bibinfo {author} {\bibfnamefont {Alain}\ \bibnamefont
  {Barrat}},\ }\bibfield  {title} {\enquote {\bibinfo {title} {Contact
  {Patterns} among {High} {School} {Students}},}\ }\href {\doibase
  10.1371/journal.pone.0107878} {\bibfield  {journal} {\bibinfo  {journal}
  {PLoS ONE}\ }\textbf {\bibinfo {volume} {9}},\ \bibinfo {pages} {e107878}
  (\bibinfo {year} {2014})}\BibitemShut {NoStop}%
\bibitem [{\citenamefont {Strelioff}\ \emph {et~al.}(2007)\citenamefont
  {Strelioff}, \citenamefont {Crutchfield},\ and\ \citenamefont
  {Hübler}}]{strelioff_inferring_2007}%
  \BibitemOpen
  \bibfield  {author} {\bibinfo {author} {\bibfnamefont {Christopher~C.}\
  \bibnamefont {Strelioff}}, \bibinfo {author} {\bibfnamefont {James~P.}\
  \bibnamefont {Crutchfield}}, \ and\ \bibinfo {author} {\bibfnamefont
  {Alfred~W.}\ \bibnamefont {Hübler}},\ }\bibfield  {title} {\enquote
  {\bibinfo {title} {Inferring {Markov} chains: {Bayesian} estimation, model
  comparison, entropy rate, and out-of-class modeling},}\ }\href {\doibase
  10.1103/PhysRevE.76.011106} {\bibfield  {journal} {\bibinfo  {journal}
  {Physical Review E}\ }\textbf {\bibinfo {volume} {76}},\ \bibinfo {pages}
  {011106} (\bibinfo {year} {2007})}\BibitemShut {NoStop}%
\bibitem [{\citenamefont {Polansky}(2007)}]{polansky_detecting_2007}%
  \BibitemOpen
  \bibfield  {author} {\bibinfo {author} {\bibfnamefont {Alan~M.}\ \bibnamefont
  {Polansky}},\ }\bibfield  {title} {\enquote {\bibinfo {title} {Detecting
  change-points in {Markov} chains},}\ }\href {\doibase
  10.1016/j.csda.2006.11.040} {\bibfield  {journal} {\bibinfo  {journal}
  {Computational Statistics \& Data Analysis}\ }\textbf {\bibinfo {volume}
  {51}},\ \bibinfo {pages} {6013--6026} (\bibinfo {year} {2007})}\BibitemShut
  {NoStop}%
\bibitem [{\citenamefont {Arnesen}\ \emph {et~al.}(2016)\citenamefont
  {Arnesen}, \citenamefont {Holsclaw},\ and\ \citenamefont
  {Smyth}}]{arnesen_bayesian_2016}%
  \BibitemOpen
  \bibfield  {author} {\bibinfo {author} {\bibfnamefont {Petter}\ \bibnamefont
  {Arnesen}}, \bibinfo {author} {\bibfnamefont {Tracy}\ \bibnamefont
  {Holsclaw}}, \ and\ \bibinfo {author} {\bibfnamefont {Padhraic}\ \bibnamefont
  {Smyth}},\ }\bibfield  {title} {\enquote {\bibinfo {title} {Bayesian
  {Detection} of {Changepoints} in {Finite}-{State} {Markov} {Chains} for
  {Multiple} {Sequences}},}\ }\href {\doibase 10.1080/00401706.2015.1044118}
  {\bibfield  {journal} {\bibinfo  {journal} {Technometrics}\ }\textbf
  {\bibinfo {volume} {58}},\ \bibinfo {pages} {205--213} (\bibinfo {year}
  {2016})}\BibitemShut {NoStop}%
\bibitem [{\citenamefont {Metropolis}\ \emph {et~al.}(1953)\citenamefont
  {Metropolis}, \citenamefont {Rosenbluth}, \citenamefont {Rosenbluth},
  \citenamefont {Teller},\ and\ \citenamefont
  {Teller}}]{metropolis_equation_1953}%
  \BibitemOpen
  \bibfield  {author} {\bibinfo {author} {\bibfnamefont {Nicholas}\
  \bibnamefont {Metropolis}}, \bibinfo {author} {\bibfnamefont {Arianna~W.}\
  \bibnamefont {Rosenbluth}}, \bibinfo {author} {\bibfnamefont {Marshall~N.}\
  \bibnamefont {Rosenbluth}}, \bibinfo {author} {\bibfnamefont {Augusta~H.}\
  \bibnamefont {Teller}}, \ and\ \bibinfo {author} {\bibfnamefont {Edward}\
  \bibnamefont {Teller}},\ }\bibfield  {title} {\enquote {\bibinfo {title}
  {Equation of {State} {Calculations} by {Fast} {Computing} {Machines}},}\
  }\href {\doibase 10.1063/1.1699114} {\bibfield  {journal} {\bibinfo
  {journal} {The Journal of Chemical Physics}\ }\textbf {\bibinfo {volume}
  {21}},\ \bibinfo {pages} {1087} (\bibinfo {year} {1953})}\BibitemShut
  {NoStop}%
\bibitem [{\citenamefont {Hastings}(1970)}]{hastings_monte_1970}%
  \BibitemOpen
  \bibfield  {author} {\bibinfo {author} {\bibfnamefont {W.~K.}\ \bibnamefont
  {Hastings}},\ }\bibfield  {title} {\enquote {\bibinfo {title} {Monte {Carlo}
  sampling methods using {Markov} chains and their applications},}\ }\href
  {\doibase 10.1093/biomet/57.1.97} {\bibfield  {journal} {\bibinfo  {journal}
  {Biometrika}\ }\textbf {\bibinfo {volume} {57}},\ \bibinfo {pages} {97 --109}
  (\bibinfo {year} {1970})}\BibitemShut {NoStop}%
\bibitem [{\citenamefont {Peixoto}(2017)}]{peixoto_nonparametric_2017}%
  \BibitemOpen
  \bibfield  {author} {\bibinfo {author} {\bibfnamefont {Tiago~P.}\
  \bibnamefont {Peixoto}},\ }\bibfield  {title} {\enquote {\bibinfo {title}
  {Nonparametric {Bayesian} inference of the microcanonical stochastic block
  model},}\ }\href {\doibase 10.1103/PhysRevE.95.012317} {\bibfield  {journal}
  {\bibinfo  {journal} {Physical Review E}\ }\textbf {\bibinfo {volume} {95}},\
  \bibinfo {pages} {012317} (\bibinfo {year} {2017})}\BibitemShut {NoStop}%
\bibitem [{\citenamefont {Karsai}\ \emph {et~al.}(2017)\citenamefont {Karsai},
  \citenamefont {Jo},\ and\ \citenamefont {Kaski}}]{karsai2017bursty}%
  \BibitemOpen
  \bibfield  {author} {\bibinfo {author} {\bibfnamefont {M{\'a}rton}\
  \bibnamefont {Karsai}}, \bibinfo {author} {\bibfnamefont {Hang-Hyun}\
  \bibnamefont {Jo}}, \ and\ \bibinfo {author} {\bibfnamefont {Kimmo}\
  \bibnamefont {Kaski}},\ }\href@noop {} {\enquote {\bibinfo {title} {Bursty
  human dynamics},}\ } (\bibinfo {year} {2017})\BibitemShut {NoStop}%
\bibitem [{\citenamefont {Masuda}\ \emph {et~al.}(2013)\citenamefont {Masuda},
  \citenamefont {Klemm},\ and\ \citenamefont
  {Eguíluz}}]{masuda_temporal_2013}%
  \BibitemOpen
  \bibfield  {author} {\bibinfo {author} {\bibfnamefont {Naoki}\ \bibnamefont
  {Masuda}}, \bibinfo {author} {\bibfnamefont {Konstantin}\ \bibnamefont
  {Klemm}}, \ and\ \bibinfo {author} {\bibfnamefont {Víctor~M.}\ \bibnamefont
  {Eguíluz}},\ }\bibfield  {title} {\enquote {\bibinfo {title} {Temporal
  {Networks}: {Slowing} {Down} {Diffusion} by {Long} {Lasting}
  {Interactions}},}\ }\href {\doibase 10.1103/PhysRevLett.111.188701}
  {\bibfield  {journal} {\bibinfo  {journal} {Physical Review Letters}\
  }\textbf {\bibinfo {volume} {111}},\ \bibinfo {pages} {188701} (\bibinfo
  {year} {2013})}\BibitemShut {NoStop}%
\end{thebibliography}%

\onecolumngrid
\pagebreak
\appendix
\section{Other datasets}\label{app:datasets}

\begin{figure*}\centering
  \begin{tabular}{cc}
    \begin{minipage}{.45\textwidth}\centering
      (I) Primary school\\
      \begin{overpic}[width=\columnwidth]{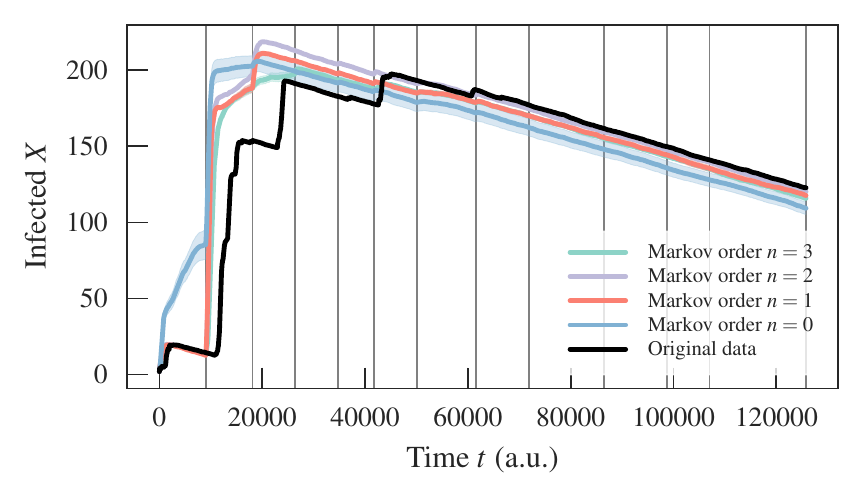}\put(0,5){(a)}\end{overpic}\\
      \begin{overpic}[width=\columnwidth]{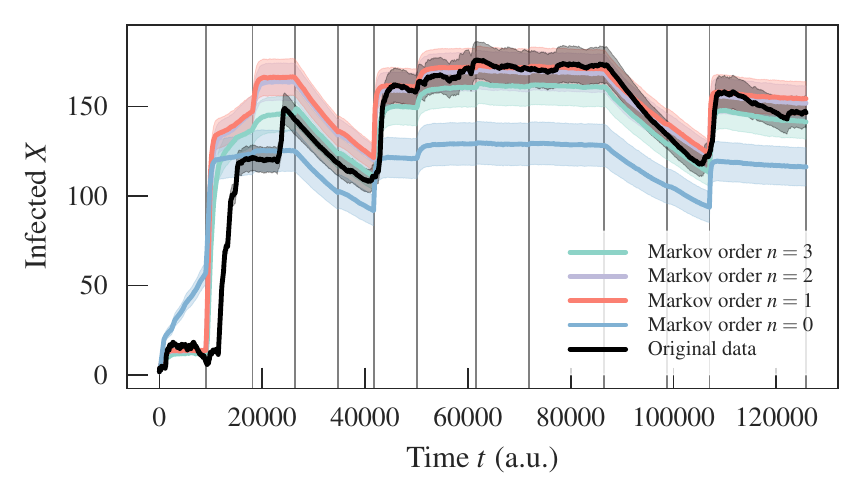}\put(0,5){(b)}\end{overpic}
      \includegraphics[width=.24\columnwidth]{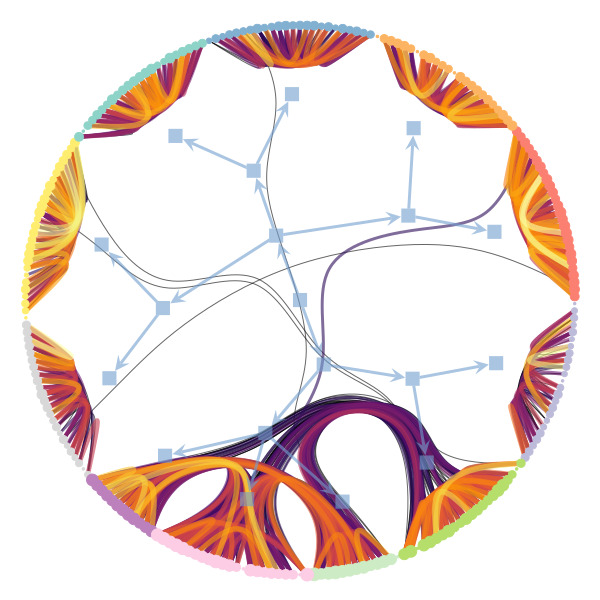}
      \includegraphics[width=.24\columnwidth]{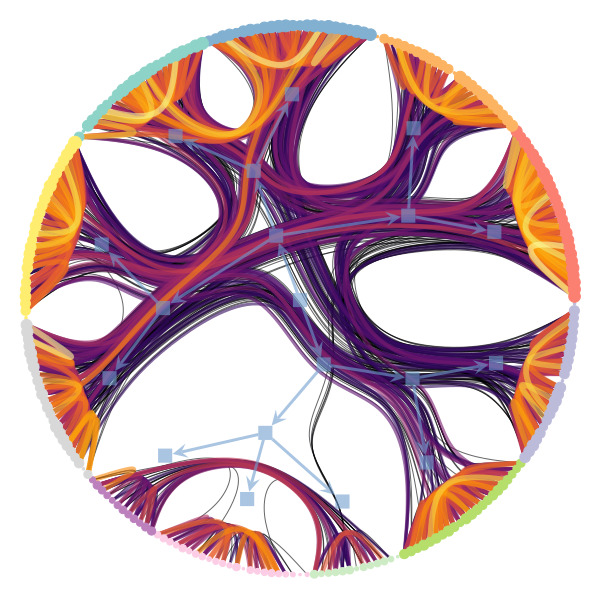}
      \includegraphics[width=.24\columnwidth]{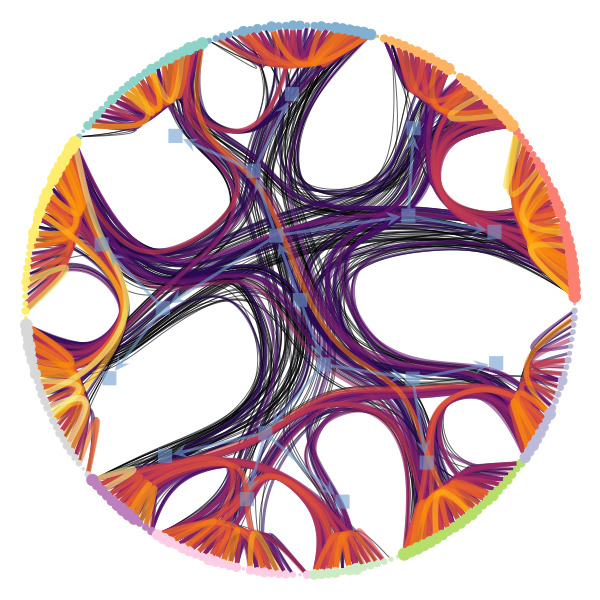}
      \includegraphics[width=.24\columnwidth]{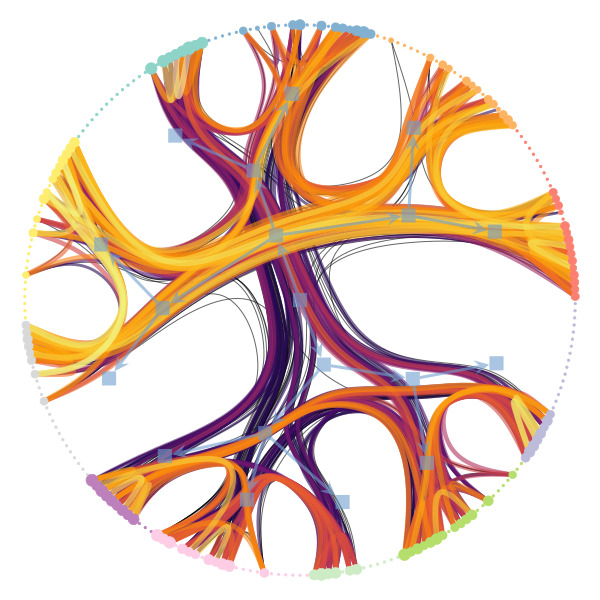}\\
      \includegraphics[width=.24\columnwidth]{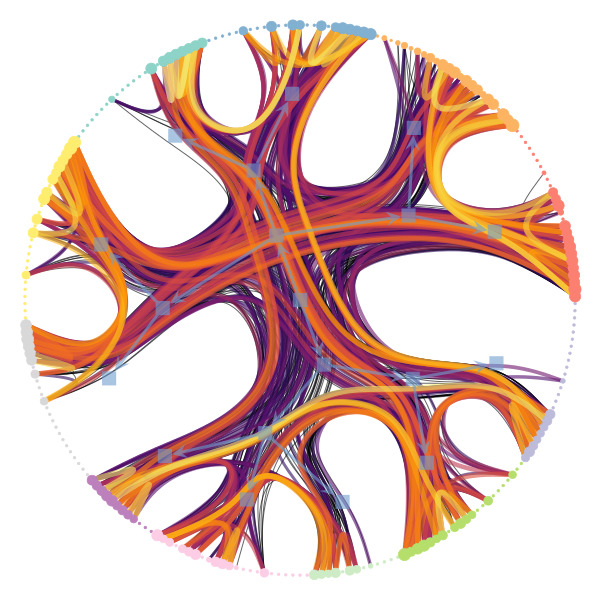}
      \includegraphics[width=.24\columnwidth]{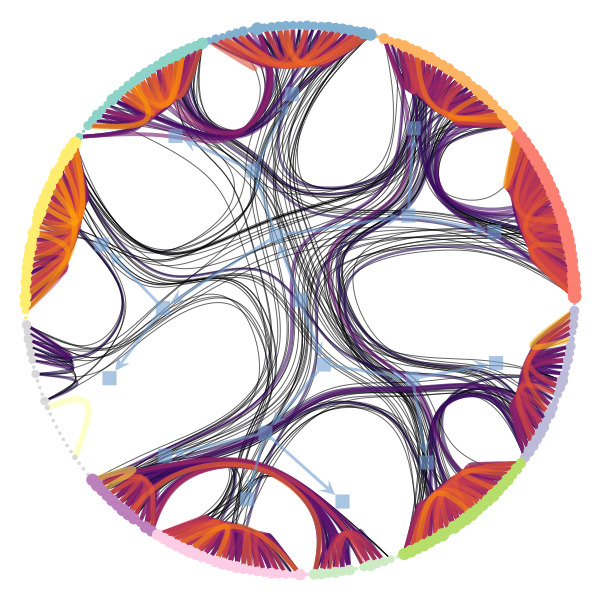}
      \includegraphics[width=.24\columnwidth]{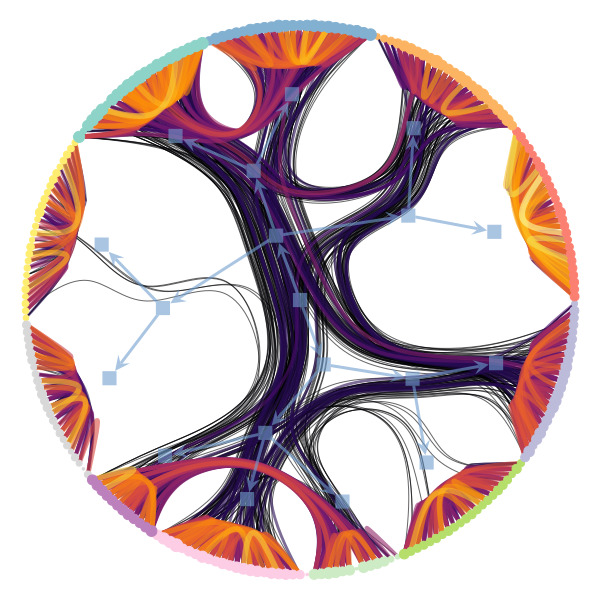}
      \includegraphics[width=.24\columnwidth]{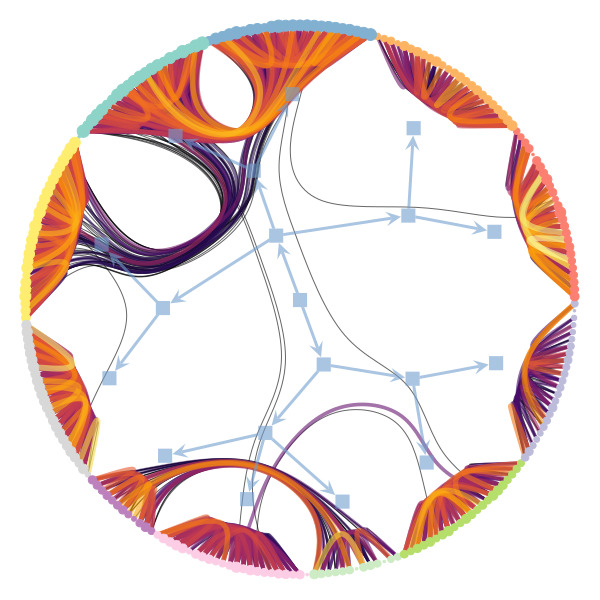}
    \end{minipage}
    \hspace{1em}&\hspace{1em}
    \begin{minipage}{.45\textwidth}
      (II) Hospital\\
      \begin{overpic}[width=\columnwidth]{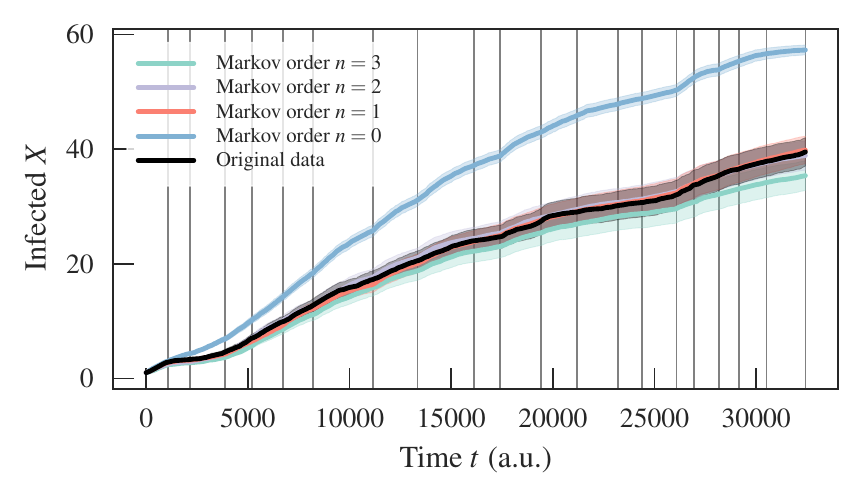}\put(0,5){(a)}\end{overpic}\\
      \begin{overpic}[width=\columnwidth]{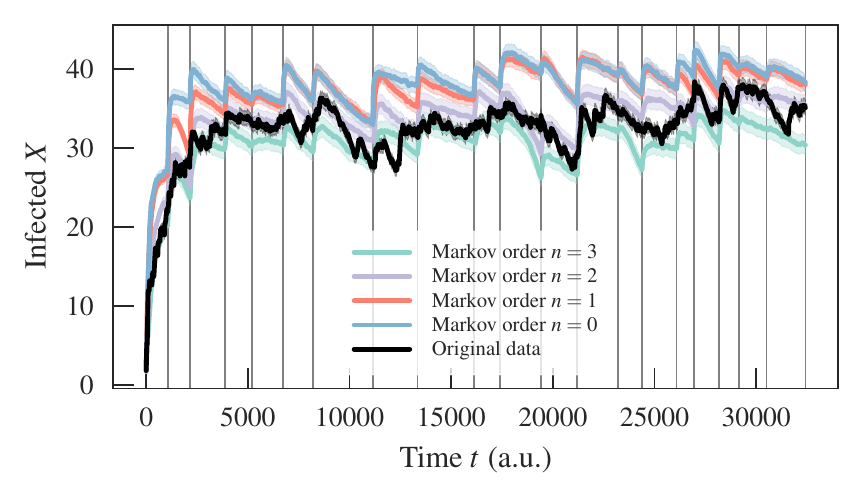}\put(0,5){(b)}\end{overpic}
      \includegraphics[width=.24\columnwidth]{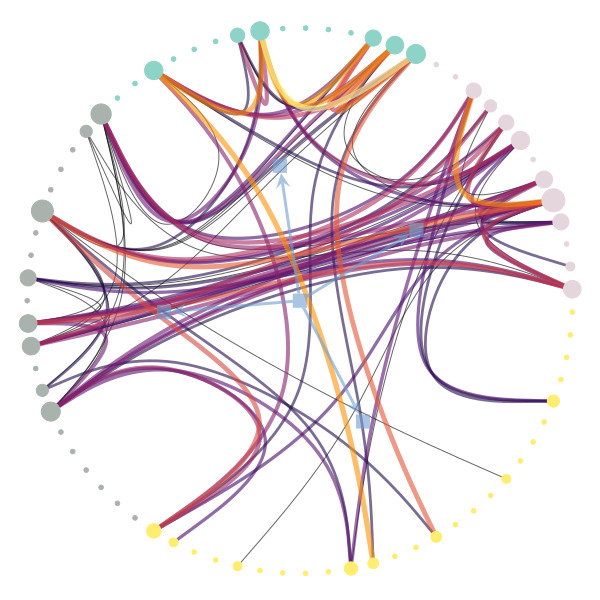}
      \includegraphics[width=.24\columnwidth]{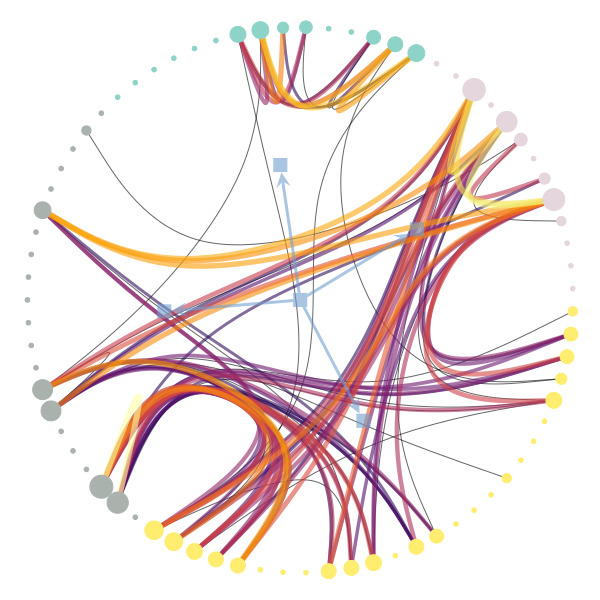}
      \includegraphics[width=.24\columnwidth]{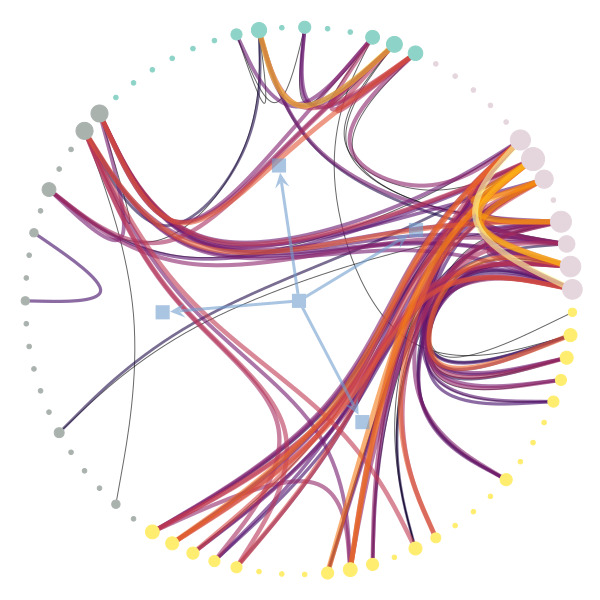}
      \includegraphics[width=.24\columnwidth]{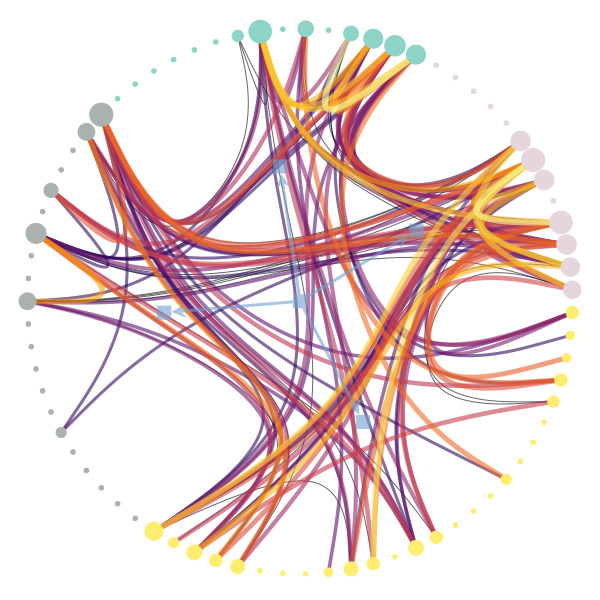}\\
      \includegraphics[width=.24\columnwidth]{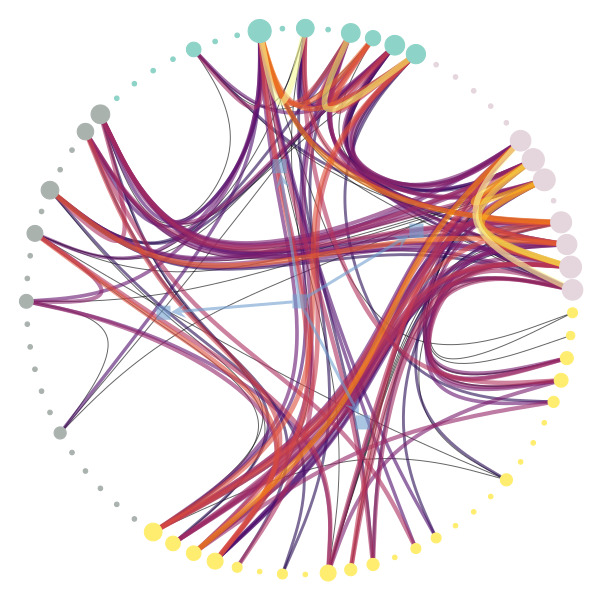}
      \includegraphics[width=.24\columnwidth]{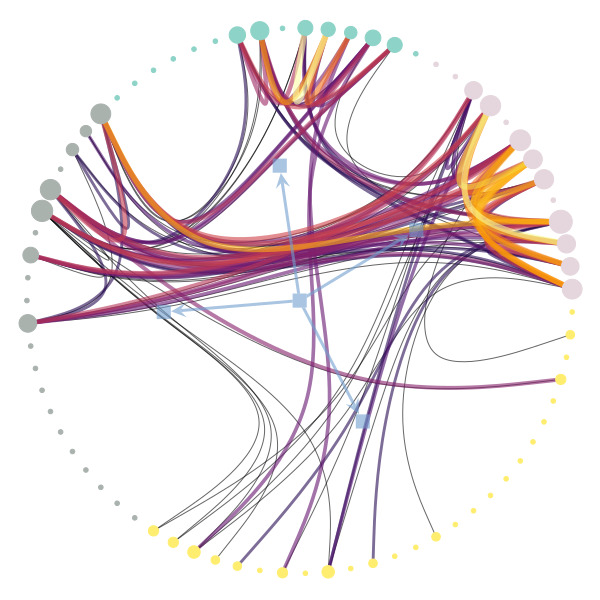}
      \includegraphics[width=.24\columnwidth]{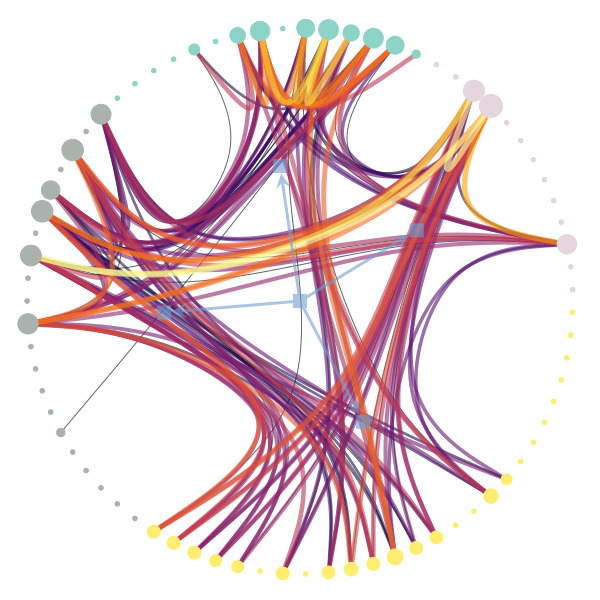}
      \includegraphics[width=.24\columnwidth]{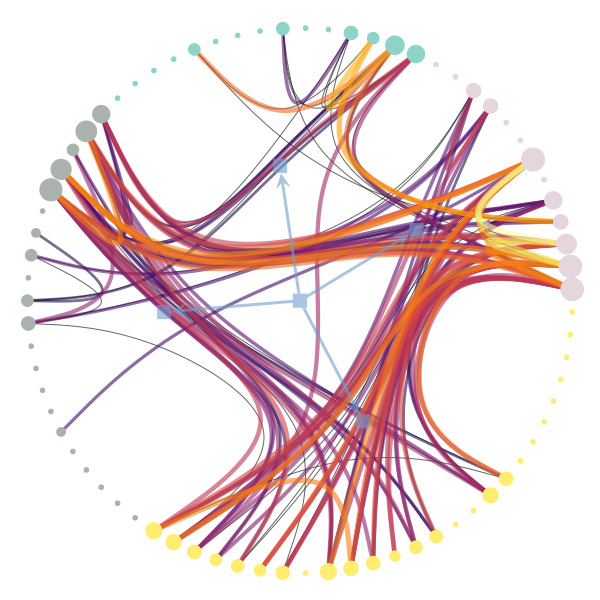}
  \end{minipage}
  \end{tabular}
  \caption{(Above) Number of infected nodes over time $X(t)$ for
  temporal networks between (I) students in a primary
  school~\cite{stehle_high-resolution_2011} and (II) patients and staff
  of a hospital~\cite{vanhems_estimating_2013}, considering both the
  original data and artificial time-series generated from the fitted
  nonstationary Markov model of a given order $n$, using (a) SIR [(I)
  $\beta=0.9$, $\gamma=0.001$; (II) $\beta=0.001$, $\gamma=0$] and (b)
  SIS [(I) $\beta=0.84$, $\gamma=0.01$; (II) $\beta=0.81$,
  $\gamma=0.015$] epidemic models. The vertical lines mark the position
  of the inferred change points. In all cases, the values were averaged
  over 100 independent realizations of the network model (for the
  artificial datasets) and dynamics. The shaded areas indicate the
  standard deviation of the mean. (Below) Network structure inside the
  first eight segments, as captured by a layered hierarchical
  degree-corrected stochastic block model~\cite{peixoto_inferring_2015}
  using the frequency of interactions as edge
  covariates~\cite{peixoto_nonparametric_2017} (indicated by colors),
  where each segment is considered as a different layer. The values of
  the infection and recovery rates were chosen so that the spreading
  dynamics spans the entire time range of the dataset.\label{fig:extra}}
\end{figure*}

Here we show that very similar results to those described in the main
text are also encountered for other proximity datasets. In
Fig.~\ref{fig:extra} (I) we show the analysis for the temporal behavior
of students in a primary school~\cite{stehle_high-resolution_2011},
which shows a very clear correlation of the change in infection rate and
the inferred change points. If we inspect the network structure inside
each temporal segment, we see that amounts to periods of time where the
students are either confined into classes, or mingling in larger
groups. A similar behavior is seen if Fig.~\ref{fig:extra} (II) for
people (staff and patients) in a hospital
ward~\cite{vanhems_estimating_2013}.

\end{document}